\documentclass[a4paper]{jpconf}
\usepackage{graphicx}
\usepackage{dcolumn}% Align table columns on decimal point
\usepackage{bm}% bold math

\usepackage{amsmath}
\usepackage{amsfonts}
\usepackage{amssymb}
\usepackage{hyperref}
\usepackage{subfig}

\begin{document}
\title{The Geometric Theory of Phase Transitions}

\author{Loris Di Cairano}

\address{Department of Physics and Materials Science, University of Luxembourg, L-1511 Luxembourg City, Luxembourg and}
\address{Computational Biomedicine, Institute of Neuroscience and Medicine INM-9 and Institute for Advanced Simulations IAS-5, Forschungszentrum Jülich, 52428 Jülich, Germany.}

 \ead{l.di.cairano.92@gmail.com, loris.dicairano@uni.lu}

\begin{abstract}
We develop a geometric theory of phase transitions (PTs) for Hamiltonian systems in the microcanonical ensemble. This theory allows to reformulate Bachmann's classification of PTs for finite-size systems in terms of geometric properties of the energy level sets (ELSs) associated to the Hamiltonian function. Specifically, by defining the microcanonical entropy as the logarithm of the ELS’s volume equipped with a suitable metric tensor, we obtain an exact equivalence between thermodynamics and geometry. In fact, we show that any derivative of entropy with respect to the energy variable can be associated to a specific combination of geometric curvature structures of the ELSs which, in turn, are precise combinations of the potential function derivatives. In this way, we establish a direct connection between the microscopic description provided by the Hamiltonian and the collective behavior which emerges in a PT. Finally, we also analyze the behavior of the ELSs' geometry in the thermodynamic limit, showing that non-analyticities of the energy-derivatives of the entropy are caused by non-analyticities of certain geometric properties of the ELSs around the transition point. Finally, we validate the theory studying the PTs that occur in the $\phi^4$ and Ginzburg-Landau-like models.
\end{abstract}

\paragraph*{Introduction}
Historically, phase
transitions (PTs) have been associated to non-analyticities of the derivatives of specific thermodynamics functions. In particular, Ehrenfest proposed to determine the order of a PT depending on the lowest derivative of the thermodynamic free energy which is non analytic at the transition temperature \cite{ehrenfest1933phasenumwandlungen}. A different classification, known as \textit{microcanonical
analysis}, have been proposed by Gross \cite{gross2001microcanonical,gross2005microcanonical} identifying PTs with the presence of convex region of microcanonical entropy.
Recently, Bachmann et al. \cite{bachmann2014thermodynamics,bachmann2014novel,schnabel2011microcanonical,qi2018classification,koci2017subphase,sitarachu2020exact,sitarachu2020phase} developed a novel classification of PTs called \textit{microcanonical inflection-point analysis}. The signature of a PT is represented by a least-sensitive inflection point in the derivatives of the microcanonical entropy distinguishing between \textit{independent} and \textit{dependent} PTs. Hence, an independent PT of order $2k$ ($2k-1$) occurs if there is a least-sensitive inflection point in the $2k-1$-th ($2k-2$-th) derivative of entropy and the corresponding maximum (minimum) in the $2k$-th $(2k-1$-th)
derivative of entropy is negative (positive) \cite{qi2018classification}. Although Bachmann's classification is a powerful and general criterion to characterize PTs, it does not provide any insight about the origin or possible mechanisms at the basis of the PTs. In other words, no relation between PTs and the microscopic interactions among the system's degrees of freedom (DoF) is established. 
In the late 1990s, an approach that aims to provide a deeper origin of PTs has been developed by Pettini et al. \cite{gori2018topological,franzosi2004theorem,CASETTI2000237,pettini2007geometry,PhysRevLett.79.4361,pettini2019origin}, resulting in the formulation of a topological theory of PTs. Such topological theory stems from the study of Hamiltonian dynamical systems \cite{PhysRevLett.79.4361} where, exploiting the so-called \emph{geometrization procedure}, one can identify the Hamiltonian dynamics with a geodesic flow on a Riemannian manifold defined by the accessible configuration space equipped with a suitable metric tensor. It is worth stressing that there does not exist a one-to-one relation between Riemannian manifold and Hamiltonian system; conversely, it is possible to geometrize a Hamiltonian system associating to it different metric tensors. Among all the possible Riemannian metrics, we have the Jacobi metric \cite{di2019coherent,arnol2013mathematical,pettini2007geometry,casetti1993analytic} and the Eisenhart metric \cite{eisenhart1928dynamical,di2021hamiltonian,pettini1993geometrical,casetti1996riemannian}. By numerical investigations, it has been observed that, independently by the metric tensor, geometric quantities such as scalar curvature and its fluctuations display discontinuities or cusp-like trends close to the transition energy or temperature  \cite{PhysRevLett.79.4361,CASETTI2000237,franzosi1999topological,caiani1998geometry,caiani1998hamiltonian}. These observations led us to conclude that the emergence of such a catastrophic behavior of the geometry is actually caused by a major topology change of the energy or potential level sets. Thus, this concept has been formalized in a \textit{necessity theorem} which attributes to the occurrence of PTs a topology change. Albeit it has been shown \cite{mehta2012energy,kastner2011phase} that the $\phi^{4}$-model represents a counterexample to the necessity theorem, a possible resolution has been presented in Ref. \cite{gori2018topological}. Here, the authors showed that, in the thermodynamic limit, a topology change is necessary for the occurrence of a PT.\\
At the same time, Rugh proposed a geometric approach which allows to identify the microcanonical temperature with the mean curvature of the energy level sets (ELSs) \cite{rugh2001microthermodynamic,rugh1997dynamical,rugh1998geometric}. This was the first step towards the identification of thermodynamic observables with geometric structures; however, the main flaw of this approach lies in the impossibility of identifying higher order derivatives of the entropy with well-defined geometric structures. Nevertheless, such an idea has been recently developed by Franzosi et al. in Ref. \cite{bel2020geometrical}: their results show that the differential geometry is undoubtedly a powerful and reliable tool for investigating PTs in the microcanonical ensemble. 
In a last paper \cite{di2021topology}, instead, we have shown that, adopting a revised definition of entropy as suggested by Franzosi \cite{franzosi2018microcanonical,franzosi2019microcanonical}, one can identify each derivative of the entropy with a specific geometric structure. The applicability of this definition is actually restricted to systems with a low number of DoF. \\
In this Letter, we generalize the result obtained in Ref. \cite{di2021topology} to the case of the ``standard'' Boltzmann's definition of microcanonical entropy. Through a theoretical derivation, we provide exact relations between thermodynamics and geometric properties of the ELSs. More precisely, we show that to each derivative of order $k$ of the microcanonical entropy can be associated specific combinations of geometric curvature entities such as the mean and scalar curvatures. In light of this, we show and discuss how the energy behaviors of the microcanonical entropy derivatives predicted by Bachmann find a natural explanation from a purely geometric perspective. In fact, we obtain a hierarchical collection of geometric entities such that each one of these is responsible of the occurrence of a PT of specific order $k$. Remarkably, such geometric entities are, in turn, suitable combinations of the potential function derivatives and this allows to establish a direct connection between the microscopic description provided by the Hamiltonian and the macroscopic or collective behavior which emerges in a PT. Then, we study the behavior of the geometry in the thermodynamic limit providing a conceptual relation between the Bachmann's and Ehrenfest's classifications. In this respect, we show that in the thermodynamic limit, the geometric curvature entities must develop a discontinuity in correspondence of the maximum or minimum point arising in the derivatives of entropy at finite $n$ according with Bachmann \cite{qi2018classification}. Finally, as a proof-of-concept, we validate the theory applying the mathematical tools developed in this Letter to two Hamiltonian systems, namely, the $\phi^4$-model and the Ginzburg-Landau-like model.

%========== Mathematical Background =======

\paragraph{Mathematical Background.}
Let us consider a generic autonomous Hamiltonian system, described by the Hamiltonian function $H:\Lambda\subset\mathbb{R}^{2N}\rightarrow\mathbb{R}$ where $\Lambda$ is the phase space and $H(\bm{x})=E$ is the energy value associated to the representative point, $\bm{x}=\{\bm{p},\bm{q}\}\subset\Lambda$, of the system. By fixing a specific energy value, the dynamics of the representative point of the system lies on the ELS \cite{rousset2010free,pettini2007geometry}:
\begin{equation}\label{def:energy_level_sets}
    \Sigma^{H}_{E}:=\{\bm{x}\in \Lambda\,|\, H(\bm{x})=E\}\subset\mathbb{R}^{n}.
\end{equation} 
Thus, the Boltzmann's definition of entropy which is given by $(n=2N)$:
\begin{equation}\label{def:boltzmann_entropy}
    S(E):=\log\int_{\Lambda}\delta(E-H(\bm{x}))~d^{n}\bm{x},
\end{equation}
where $\delta$ is the Dirac delta function, can be reinterpreted in a geometric framework introducing a well-precise metric tensor, $g_{\Lambda}$. Thus, the entropy function coincides with the logarithm of the Riemannian volume of the ELS, $\Sigma_{E}^{H}$, i.e.:
% \begin{equation}\label{def:riemannian_volume}
%     vol^{g_{\Lambda}}(\Sigma_{E}^{H})=\int_{\Sigma_{E}^{H}}~d\mu^{g_{\Lambda}}_{\Sigma^{H}_{E}}(\bm{x}),
% \end{equation} 
% where 
% which allows to define the entropy function as below
\begin{equation}\label{def:entropy_geometry}
    S_{g_{\Lambda}}(E):=\log\left(vol^{g_{\Lambda}}(\Sigma^{H}_{E})\right),
\end{equation}
In order to define the metric tensor, $g_{\Lambda}$, it is necessary to introduce a few further concepts. Let
us consider the range of all possible energies accessible to the system, $\mathcal{E}:= [E_0,E_1]\subset\mathbb{R}$. 
Thus, for any energy value, $E\in\mathcal{E}$, there exists an ELS, $\Sigma_E^H$, which is a $n-1$-dimensional hypersurface embedded in $\mathbb{R}^{n}$. Hence, we can define the collection $\{\Sigma_{E}^{H}\}_{E\in\mathcal{E}}$, i.e., a phase space foliation defined by $\Lambda=\bigcup_{E\in\mathcal{E}}\;\Sigma_{E}^{H}$.
Therefore, we introduce a curvilinear coordinate system on $\Lambda$, i.e., $\{u^{\alpha}\}_{\alpha=0}^{n-1}$, such that $u^{0}=E$ and $\{u^{i}\}_{i=1}^{n-1}$ is the system of coordinates on $\Sigma_{E}^{H}$ together with the vector basis $\{\partial_{u^{\alpha}}\}_{\alpha=0}^{n-1}$ and its dual $\{du^{\alpha}\}_{\alpha=0}^{n-1}$ such that $du^{\alpha}(\partial_{u^{\beta}})=\delta^{\alpha}_{\beta}$.\\
In so doing, a Riemannian metric tensor is defined on the whole phase space \cite{zhou2013simple}:
\begin{equation}\label{def:metric_tensor_zhou}
    g_{\Lambda}=\chi^{2}du^{0}\otimes du^{0}+h_{ij}du^{i}\otimes du^{j}
\end{equation}
where $\chi:=1/\|\nabla H\|$ and $\nabla$ is the gradient operator defined with respect to the Euclidean metric and $h_{ij}$ is the Euclidean metric induced on the ELSs.
In this framework, the volume measure of the phase space (equivalent to the Liouville measure) is $ d\eta^{g_{\Lambda}}=d\sigma_{\Sigma_{E}^{H}}\;du^{0}/\|\nabla H\|$ where $d\sigma_{\Sigma_{E}^{H}}:=\sqrt{\det\,h}\,du^{1}\ldots du^{n-1}$, and, thereby, the induced measure on the ELS is:
\begin{equation}
    d\mu^{g_{\Lambda}}_{\Sigma_{E}^{H}}:=d\eta^{g_{\Lambda}}\bigg|_{\Sigma_{E}^{H}}=\frac{d\sigma_{\Sigma_{E}^{H}}}{\|\nabla H\|}.
\end{equation}
Thus, we obtain the definition of the $n-1$-dimensional volume of an ELS:
\begin{equation}\label{def:volume_g_Lambda}
    vol^{g_{\Lambda}}(\Sigma_{E}^{H}):=\int_{\Sigma_{E}^{H}}d\mu^{g_{\Lambda}}_{\Sigma_{E}^{H}},
\end{equation}
which leads to the definition of entropy in Eq. \eqref{def:entropy_geometry}. 
It is worth noting that a generic Hamiltonian function admit a large number of critical points \cite{matsumoto2002introduction,stein1963morse}, i.e., $\bm{x}_c\in\Lambda$, such that $\nabla H(\bm{x})|_{\bm{x}=\bm{x}_c}=\bm{0}$. Moreover, their number grows exponentially with the number of DoF \cite{casetti2009kinetic,kastner2008phase,kastner2011phase} and they can give rise to a continuous set, $\mathcal{C}\subset\Lambda$, such that $dH_{\bm{x}_c}=0$ for any $\bm{x}_c\in\mathcal{C}$. Furthermore, they can be distributed over the entire collection $\{\Sigma_{E}^{H}\}_{E\in\mathcal{E}}$. A naive argument could lead to the wrong conclusion that the critical points have a central role in a PT since, apparently, they can give rise to divergences of the ELS's volume \eqref{def:volume_g_Lambda}. In fact, due to the fact that $\chi(\bm{x})\to\infty$ for $\|\bm{x}-\bm{x}_c\|\to 0$, the ELS's volume can diverge. However, a more rigorous analysis shows that the integration measure, $d\sigma_{\Sigma_{E}^{H}}$, \emph{regularizes} the integral \eqref{def:volume_g_Lambda} also at the critical points so that the vanishing of the denominator, $\|\nabla H\|$, does not entail any divergence of the volume \cite{pettini2007geometry,nerattini2013exploring}.  % read pag 223, 241 Pettini's book.
A common assumption is to consider the Hamiltonian functions of interest to be Morse functions \cite{pettini2007geometry}, i.e., such that all its critical points are finite and isolated. This means that there exists always an ELS that does not contain any critical point. In other words, if we have $\bm{x}_c\in\Sigma_{E_*}^{H}$ with $E_*\in\mathcal{E}$, then, there exists a sufficiently large $\epsilon>0$ such that, defining $E_\epsilon:= E_{*}+\epsilon$, we have $\nabla H(\bm{x})\neq\bm{0}$ for any $\bm{x}\in\Sigma_{E_{\epsilon}}^{H}$. This allows us to safely define the metric tensor \eqref{def:metric_tensor_zhou} on $\Sigma_{E_{\epsilon}}^{H}$ and to study the ELSs' evolution, $\Sigma_{E_{\epsilon}}^{H}\to\Sigma_{E^{\prime}}^{H}$, under the energy flow, $E_{\epsilon}\to E^{\prime}$.

\paragraph{Entropy Flow Equations.}  Definition \eqref{def:entropy_geometry} highlights an equivalence between thermodynamics---the microcanonical entropy on the left-hand side---and the geometry---the geometric volume of the ELSs on the right-hand side. In fact, by differentiating both sides of Eq. \eqref{def:entropy_geometry} with respect to the energy, we have a connection between the derivatives of entropy and volume. Now, the rate of expansion of the volume (first derivative) as well as the acceleration (second derivative) are approximately related to, respectively, the mean curvature \cite{rugh1997dynamical,rugh1998geometric,rugh2001microthermodynamic} and scalar curvature of the considered hypersurface \cite{gromov2019four}. It is thus natural to expect that (specific) changes of the ELSs' geometry, which can be observed by means of energy variations of volume, can cause changes of the thermodynamic properties of the associated physical system. In other words, we expect that there exist classes of geometric observables such as the scalar and mean curvatures which are in a one-to-one correspondence with the order of the lowest entropy derivative that manifests a non-trivial behavior in energy. To formalize this concept more precisely, we compute the derivatives of the entropy function \eqref{def:entropy_geometry} with respect to $E$, namely:
\begin{equation}
        \partial_{E}S_{g_{\Lambda}}(E)=\frac{\partial_{E}vol^{g_{\Lambda}}(\Sigma^{H}_{E})}{vol^{g_{\Lambda}}(\Sigma^{H}_{E})},\label{eqn:first_derivatives}
\end{equation}
\begin{equation}
        \partial^{2}_{E}S_{g_{\Lambda}}(E)=\frac{\partial_{E}^{2}vol^{g_{\Lambda}}(\Sigma^{H}_{E})}{vol^{g_{\Lambda}}(\Sigma^{H}_{E})}-\left(\frac{\partial_{E}vol^{g_{\Lambda}}(\Sigma^{H}_{E})}{vol^{g_{\Lambda}}(\Sigma^{H}_{E})}\right)^{2},\label{eqn:second_derivatives}
\end{equation}
\begin{equation}
\begin{split}
        \partial^{3}_{E}S_{g_{\Lambda}}(E)&=\frac{\partial_{E}^{3} vol^{g_{\Lambda}}(\Sigma^{H}_{E})}{vol^{g_{\Lambda}}(\Sigma^{H}_{E})}+2\left(\frac{\partial_{E}vol^{g_{\Lambda}}(\Sigma^{H}_{E})}{vol^{g_{\Lambda}}(\Sigma^{H}_{E})}\right)^{3}\\
        &-3\frac{\partial_{E}^{2}vol^{g_{\Lambda}}(\Sigma^{H}_{E})}{vol^{g_{\Lambda}}(\Sigma^{H}_{E})}\frac{\partial_{E}vol^{g_{\Lambda}}(\Sigma^{H}_{E})}{vol^{g_{\Lambda}}(\Sigma^{H}_{E})}\label{eqn:third_derivatives}.
\end{split}
\end{equation}
It is worth stressing that the mathematical structures of the relations above hold independently by the adopted metric tensor. Finally, this computation can be easily extended to higher order derivatives of entropy.
The next step would be to show that all the terms on the right-hand sides of Eqs. \eqref{eqn:first_derivatives}, \eqref{eqn:second_derivatives} and \eqref{eqn:third_derivatives} have a purely geometric meaning and it requires to compute the volume variations.\\
However, as we already have shown (see Eq. (42) in Ref. \cite{di2021topology}), adopting the metric tensor \eqref{def:metric_tensor_zhou} the volume variations cannot be identified as averages of purely geometric observables. This is because any \textit{regular} ELS is identified by the unit normal vector field
$\bm{\nu}=\nabla H/\|\nabla H\|$
such that $g_{\Lambda}(\bm{\nu},\bm{\nu})=1$. Such a vector field encodes all the information about the geometry of the hypersurface through the Weingarten operator, that is, $\mathcal{W}_{\bm{\nu}}(\bm{X})=\nabla_{\bm{X}}\bm{\nu}$ where $\bm{X}$ is any vector on the tangent space to $\Sigma_{E}^{H}$. However, the evolution of an ELS, $\Sigma_{E}^{H}\to\Sigma_{E^{\prime}}^{H}$, upon energy variation, $E\to E^{\prime}$, is generated by the vector field $\partial_{u^{0}}$ such that $dH(\partial_{u^{0}})=1$ \cite{hirsch2012differential}. In particular, it is easy to show that $(u^0\equiv E)$ \cite{di2021topology}:
\begin{equation}\label{def:zeta_vector}
    \bm{\zeta}:=\partial_{E}=\chi\bm{\nu}.
\end{equation}
We note that the vector $\bm{\zeta}$ is not normalized with respect to the metric tensor \eqref{def:metric_tensor_zhou}, i.e., $g_{\Lambda}(\bm{\zeta},\bm{\zeta})=\chi^{-2}$. In fact, since $\bm{\zeta}$ generates the diffeomorphism between ELSs, the first volume variation is given by the Lie derivative with respect to $\bm{\zeta}$ of the volume form, i.e.
\begin{equation}\label{def:first_variation_zeta}
    \begin{split}
        \partial_{E}vol^{g_{\Lambda}}(\Sigma^{H}_{E})=\int_{\Sigma_{E}^{H}}\mathcal{L}_{\bm{\zeta}}\left(\chi~d\sigma^{g_{\Lambda}}_{\Sigma_{E}^{H}}\right).
    \end{split}
\end{equation}
By exploiting $\mathcal{L}_{\bm{\zeta}}=\chi\mathcal{L}_{\bm{\nu}}$ and knowing that \cite{zhou2013simple} 
\begin{equation}\label{eqn:lie_der_dsigma}
    \mathcal{L}_{\bm{\nu}}d\sigma_{\Sigma_{E}^{H}}^{g_{\Lambda}}=Tr[\mathcal{W}_{\bm{\nu}}]d\sigma_{\Sigma_{E}^{H}}^{g_{\Lambda}},
\end{equation}
where $Tr[\mathcal{W}_{\bm{\nu}}]=div(\bm{\nu})$ (see Eq. (S.15) in Supplementary Information (SI)), we get
\begin{equation}\label{eqn:mean_curv}
\begin{split}
    \mathcal{L}_{\bm{\zeta}}\left(\chi d\sigma^{g_{\Lambda}}_{\Sigma_{E}^{H}}\right)&=\left[\chi\; Tr[\mathcal{W}_{\bm{\nu}}]+\mathcal{L}_{\bm{\nu}}\chi\right]\chi~d\sigma^{g_{\Lambda}}_{\Sigma_{E}^{H}}.
\end{split}
\end{equation}
Therefore, apart from $Tr[\mathcal{W}_{\bm{\nu}}]$ which is a geometric observable, a further term $(\mathcal{L}_{\bm{\nu}}\chi)$ appears and it cannot be identified with any geometric object. In analogy with Eq. \eqref{eqn:lie_der_dsigma}, in order to have a fully geometric description of the volume variation, one should interpret the whole right-hand side of Eq. \eqref{eqn:mean_curv} as the trace of a new Weingarten operator defined by $\bm{\zeta}$. However, it is evident that, being $\bm{\zeta}$ not normalized, the latter cannot be adopted.\\
Nevertheless, we can overcome this apparent limitation by introducing a \textit{conformal-like transformation} of the metric tensor \eqref{def:metric_tensor_zhou} which does not change the physical description of the problem. In fact, since all the thermodynamic properties of a physical system are encoded in the entropy function \eqref{def:metric_tensor_zhou}, one can find another metric tensor, $\widetilde{g}$, which gives rise to the same thermodynamic description of the physical system provided that $d\mu^{\widetilde{g}}=d\mu^{g_{\Lambda}}$.\\
Therefore, as already shown by Gori \cite{gori2018topological,gori_thesis}, by performing a change of coordinates such that:
\begin{equation}\label{def:rescaling_metric}
    dx^{0}=\chi~du^{0},\quad dx^{i}=\chi^{-\frac{1}{n-1}}~du^{i},\qquad \forall~i\in[1,n],
\end{equation}
whose components of the metric tensor are \cite{gori2018topological,gori_thesis}
\begin{equation}\label{def:rescaling_components_metric}
    \widetilde{g}_{00}=\chi^{-2}g_{00},\quad \widetilde{h}_{ij}=\chi^{\frac{2}{n-1}}~h_{ij},
\end{equation}
we get an equivalent metric to \eqref{def:metric_tensor_zhou} defined by
\begin{equation}\label{def:conformal_metric}
    \widetilde{g}=dx^{0}\otimes dx^{0}+\widetilde{h}_{ij}~dx^{i}\otimes dx^{j},
\end{equation}
and the physical description is preserved in the following sense. The rescaling in Eq. \eqref{def:rescaling_components_metric} has the property to conserve the Riemannian volume form:
\begin{equation}
\begin{split}
    dvol^{\widetilde{g}}&=(det~\widetilde{h})^{1/2} dx^{0}dx^{1}\ldots dx^{n}\\
    &=\chi(det~h)^{1/2} du^{0}du^{1}\ldots du^{n}=dvol^{g_{\Lambda}},
\end{split}
\end{equation}
as well as the Riemannian area form:
\begin{equation}
\begin{split}
    d\eta_{\Sigma_{E}^{H}}^{\widetilde{g}}&=(det~\widetilde{h})^{1/2} dx^{1}\ldots dx^{n}\\
    &=\chi(det~h)^{1/2} du^{1}\ldots du^{n}=d\mu_{\Sigma_{E}^{H}}^{g_{\Lambda}}.
\end{split}
\end{equation}
Hence, the definition of entropy given in Eq. \eqref{def:entropy_geometry} reduced to the following (equivalent) form:
\begin{equation}\label{def:entropy_gori}
    S_{\widetilde{g}}(E):=\log\int_{\Sigma_{E}^{H}}d\eta^{\widetilde{g}}_{\Sigma_{E}^{H}}.
\end{equation}
However, the vector field $\bm{\zeta}$ is now normalized, i.e.:
\begin{equation}
    \widetilde{g}(\bm{\zeta},\bm{\zeta})=\chi^{-2}g_{\Lambda}(\bm{\zeta},\bm{\zeta})=1.
\end{equation}
Therefore, we make the identification
\begin{equation}\label{def:E_derivative}
    \partial_{E}\equiv\partial_{x^{0}}:=\bm{\xi},
\end{equation}
where $\bm{\xi}$ is also the unit normal vector to the ELSs.
In this setting, we can define the Weingarten operator, regarded as a tensor field of order 1 covariant and 1 controvariant, i.e.:
\begin{equation}\label{def:weingarten_new}
    \mathcal{W}^{\widetilde{g}}_{\bm{\xi}}=\chi~\mathcal{W}_{\bm{\nu}}+\chi^{-1}\partial_{E}\chi\frac{1\!\!1_{\Sigma_{E}}}{n-1},
\end{equation}
where $\mathcal{W}_{\nu}$ is the Weingarten operator defined in Eq. \eqref{eqn:lie_der_dsigma} whereas $1\!\!1_{\Sigma_{E}}:=(h^{\widetilde{g}})^{-1}h^{\widetilde{g}}$ is the identity operator on the tangent space to the ELS $\Sigma_{E}^{H}$ (see also SI for further details). In summary, the description provided by metric tensor, $g_{\Lambda}$, does not allow to identify the volume variations with purely geometric entities; in order to achieve this purpose, we exploited a conformal-like transformation \eqref{def:rescaling_components_metric} of the metric tensor $g_{\Lambda}$ which does not affect the thermodynamic description and we obtained a new metric tensor $\widetilde{g}$ \eqref{def:conformal_metric} which realizes this identification. 
It is easy to show that the trace of the new Weingarten operator \eqref{def:weingarten_new} coincides with the right-hand side of Eq. \eqref{eqn:mean_curv}, i.e.,
$Tr[\mathcal{W}^{\widetilde{g}}_{\bm{\xi}}]=\chi Tr[\mathcal{W}_{\bm{\nu}}]+\mathcal{L}_{\bm{\nu}}\chi$. Hence, the first variation of volume defined in Eq. \eqref{def:entropy_gori} coincides with the geometric average of the trace of the new Weingarten operator:
\begin{equation}\label{eqn:first_variation_volume}
    \partial_{E}vol^{\widetilde{g}}(\Sigma_{E}^{H})=\int_{\Sigma_{E}^{H}}Tr^{\widetilde{g}}[\mathcal{W}^{\widetilde{g}}_{\bm{\xi}}]~d\eta^{\widetilde{g}}_{\Sigma_{E}^{H}}.
\end{equation}
Note that this is proportional to the mean curvature of $\Sigma_{E}^{H}$, i.e., $h(\Sigma_{E}^{H}):=Tr^{\widetilde{g}}[\mathcal{W}^{\widetilde{g}}_{\bm{\xi}}]/n$.
Finally, by plugging the equation above into Eq. \eqref{eqn:first_derivatives}, we get
\begin{equation}\label{eqn:entropy_trace_weingarten}
    \partial_{E}S_{\widetilde{g}}(E)=\int_{\Sigma_{E}^{H}}Tr^{\widetilde{g}}[\mathcal{W}^{\widetilde{g}}_{\bm{\xi}}]~d\rho^{\widetilde{g}}_{\Sigma_{E}^{H}},
\end{equation}
where $d\rho^{\widetilde{g}}_{\Sigma_{E}^{H}}=d\eta^{\widetilde{g}}_{\Sigma_{E}^{H}}/vol^{\widetilde{g}}(\Sigma_{E}^{H})$. As already anticipated at the beginning of this section, we notice in Eq. \eqref{eqn:entropy_trace_weingarten} that the first-order derivative of entropy is related to the geometric average of the mean curvature also called total mean curvature \cite{chen2014total}.
Before doing any step further, it is worth emphasizing the feasibility of this approach in the applications to physical systems. Although, at a first sight, Eq. \eqref{eqn:entropy_trace_weingarten} may seem only a formal relation between entropy and mean curvature, we note that the integral on the right-hand side can be easily computed, at least, numerically. In fact, the trace of the Weingarten operator reads (see Eq. (S.11) in SI):
\begin{equation}\label{def:trace_weingarten_hamiltonian}
    Tr[\mathcal{W}^{\widetilde{g}}_{\bm{\xi}}]=\frac{\Delta H}{\|\nabla H\|^{2}}-2\frac{\langle\nabla H, Hess\, H\cdot\nabla H\rangle}{\|\nabla H\|^{4}}.
\end{equation}
where $\nabla$, $\Delta$ and $\|\cdot\|$ are, respectively, the gradient, the laplacian and the norm defined with respect to the phase space-DoF. It is evident that, given any Hamiltonian function, the quantities in Eq. \eqref{def:trace_weingarten_hamiltonian} can be always computed regardless of the number of DoF. For instance, in the investigation of transitional phenomena in biological systems such as the protein folding, this approach can be employed as well provided that the environment's DoF are implicitly described. More precisely, it is necessary that the latter do not explicitly appear in the Hamiltonian function; conversely, the information about the protein-environment interactions are mimic by a protein-DoF-dependent term which can be physically interpreted as an external potential for the protein itself.

\paragraph{Second Variation of Volume.}

\begin{figure}[!htb]
    \begin{minipage}[t]{0.5\textwidth}
        \centering
        \includegraphics[width=\textwidth]{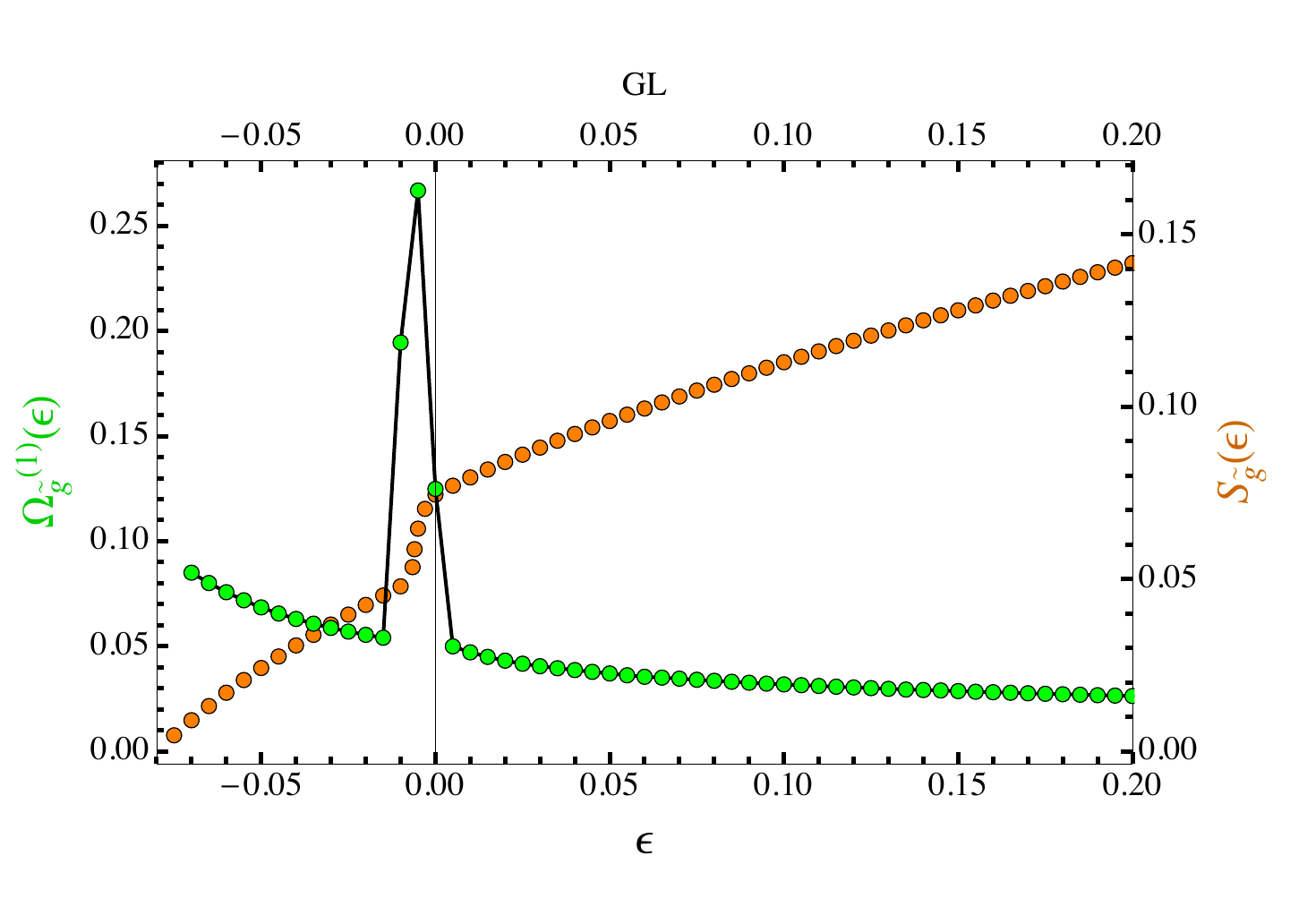}
         \caption{Plot of first-order geometric function, $\Omega_{\widetilde{g}}^{(1)}(\epsilon)$ and entropy function $S_{\widetilde{g}}(\epsilon)$ for the GL-model.}
    \label{Fig:GL_model}
    \end{minipage}
    \hfill
     \begin{minipage}[t]{0.48\textwidth}
        \centering
        \includegraphics[width=\textwidth]{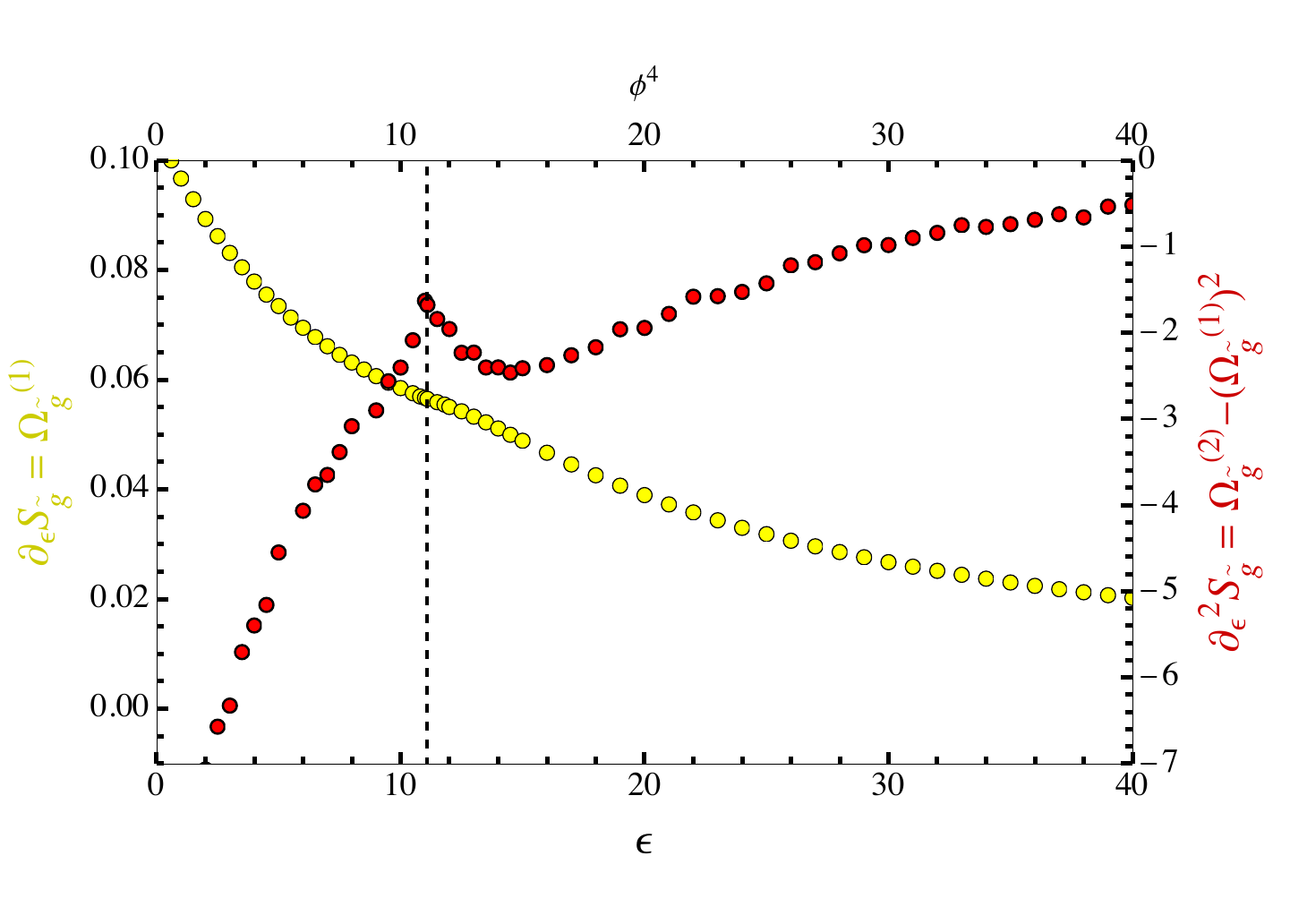}
    \caption{Plot of $\Omega_{\widetilde{g}}^{(1)}(\epsilon)$ (yellow curve)  in Eq.~\eqref{def:first_order_geometric_func} and $\partial^2_\epsilon S_{\widetilde{g}}(\epsilon)$ obtained composing the GCFs as defined in Eq. \eqref{eqn:second_derivatives_g_gori}.}
    \label{Fig:phi4_model}
    \end{minipage}
            \centering
        \includegraphics[width=7cm,height=7cm]{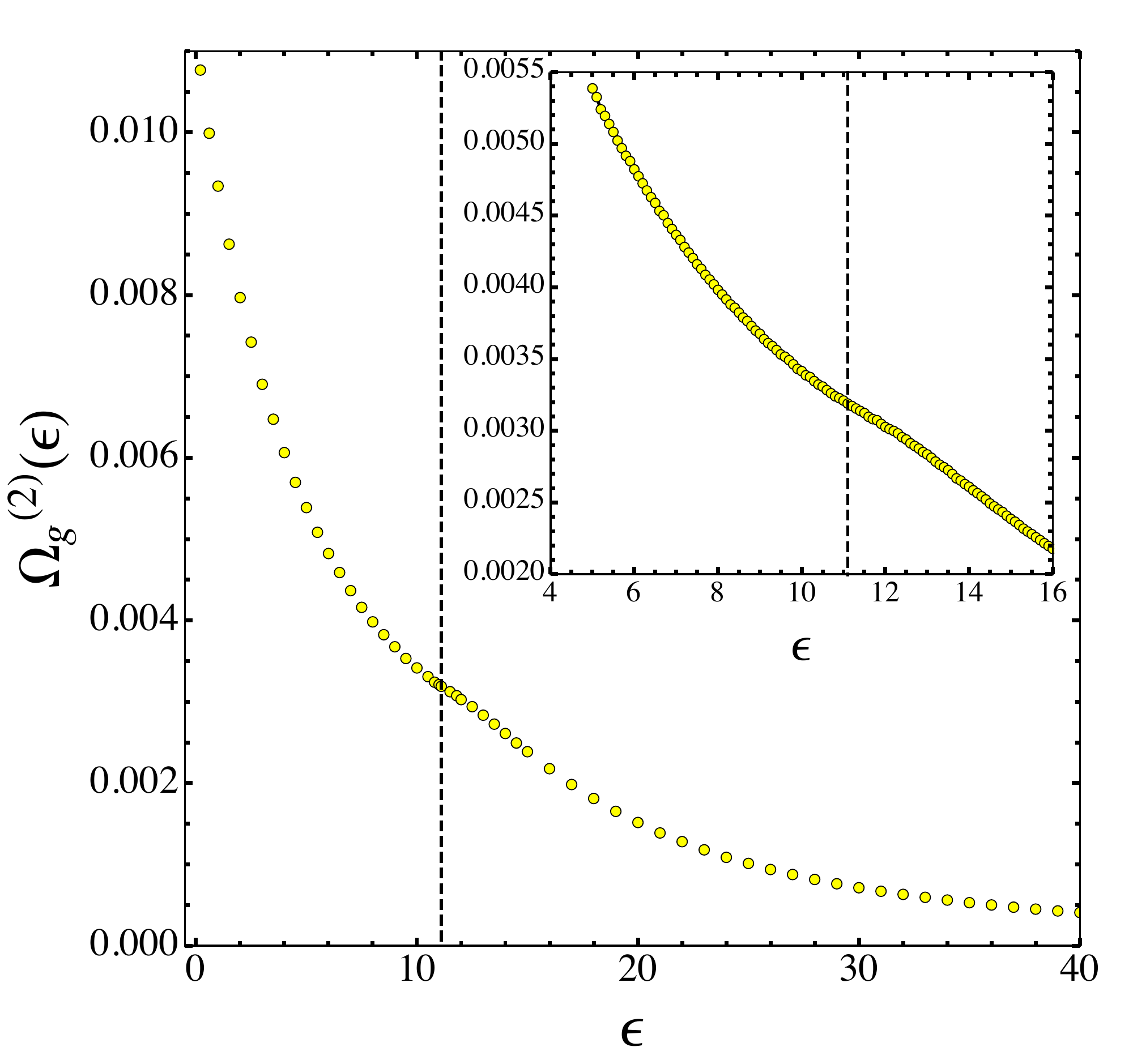}
    \caption{Plot of the second-order GCF, $\Omega_{\widetilde{g}}^{(2)}(\epsilon)$ defined in Eq. \eqref{def:second_order_geometric_func} as a function of the energy density in the $\phi^4$-model. }
    \label{Fig:phi4_omega2}
\end{figure}

\begin{figure}[!htb]
    \begin{minipage}[t]{.5\textwidth}
        \includegraphics[width=\textwidth]{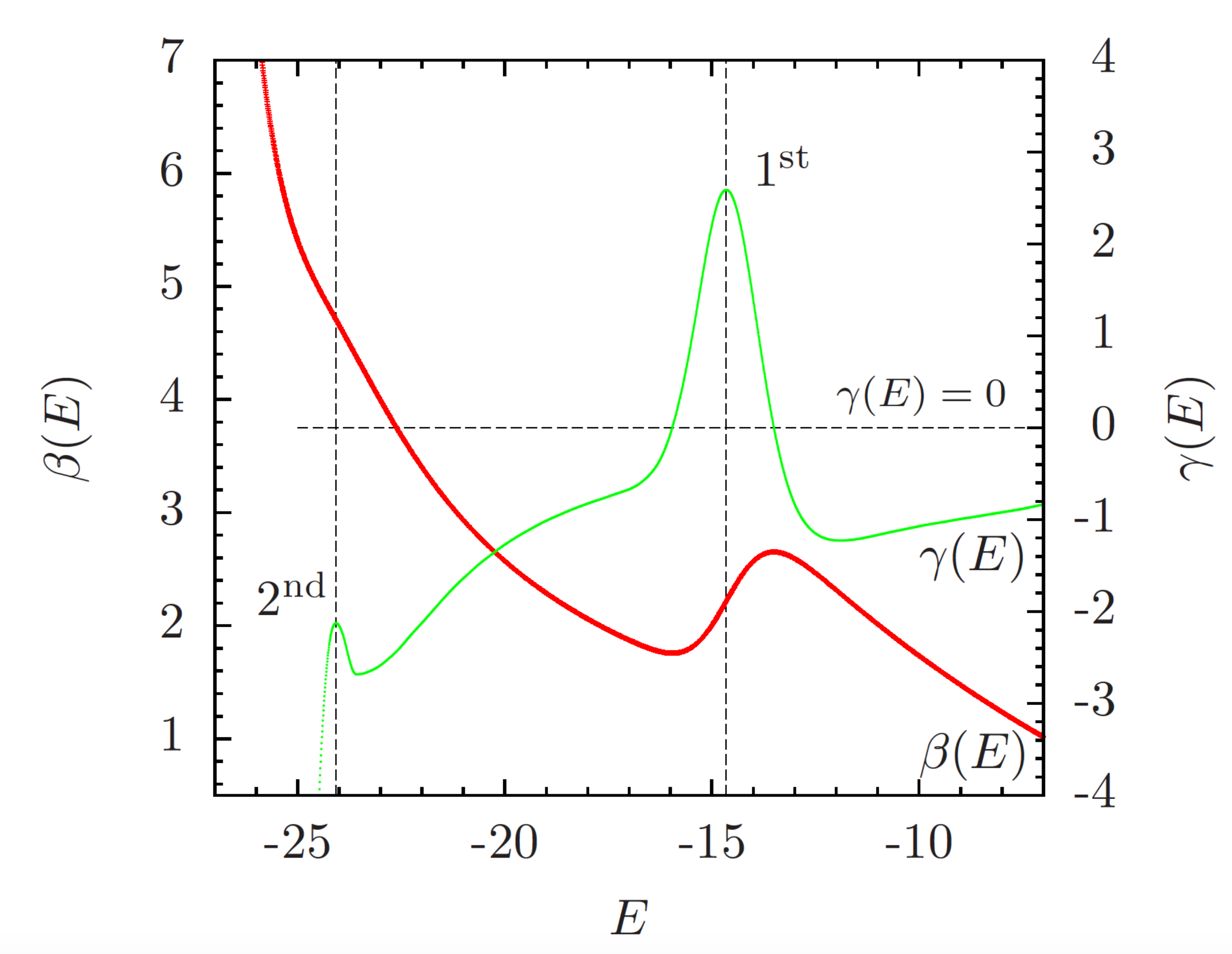}
       \caption{From Ref. \cite{koci2017subphase}. Schematic illustration of microcanonical inflection-point analysis for the inverse microcanonical temperature $\beta(E)$. The prominent back-bending region in $\beta(E)$, together with the positive-valued peak in its energy derivative $\gamma(E)$ at $E \approx-15$, indicates a first-order transition. The negative-valued peak at $E\approx -24$ corresponds to a second-order transition.}
        \label{Fig:bachmann}
    \end{minipage}
        \hfill
      \begin{minipage}[t]{.45\textwidth}
        \includegraphics[width=\textwidth]{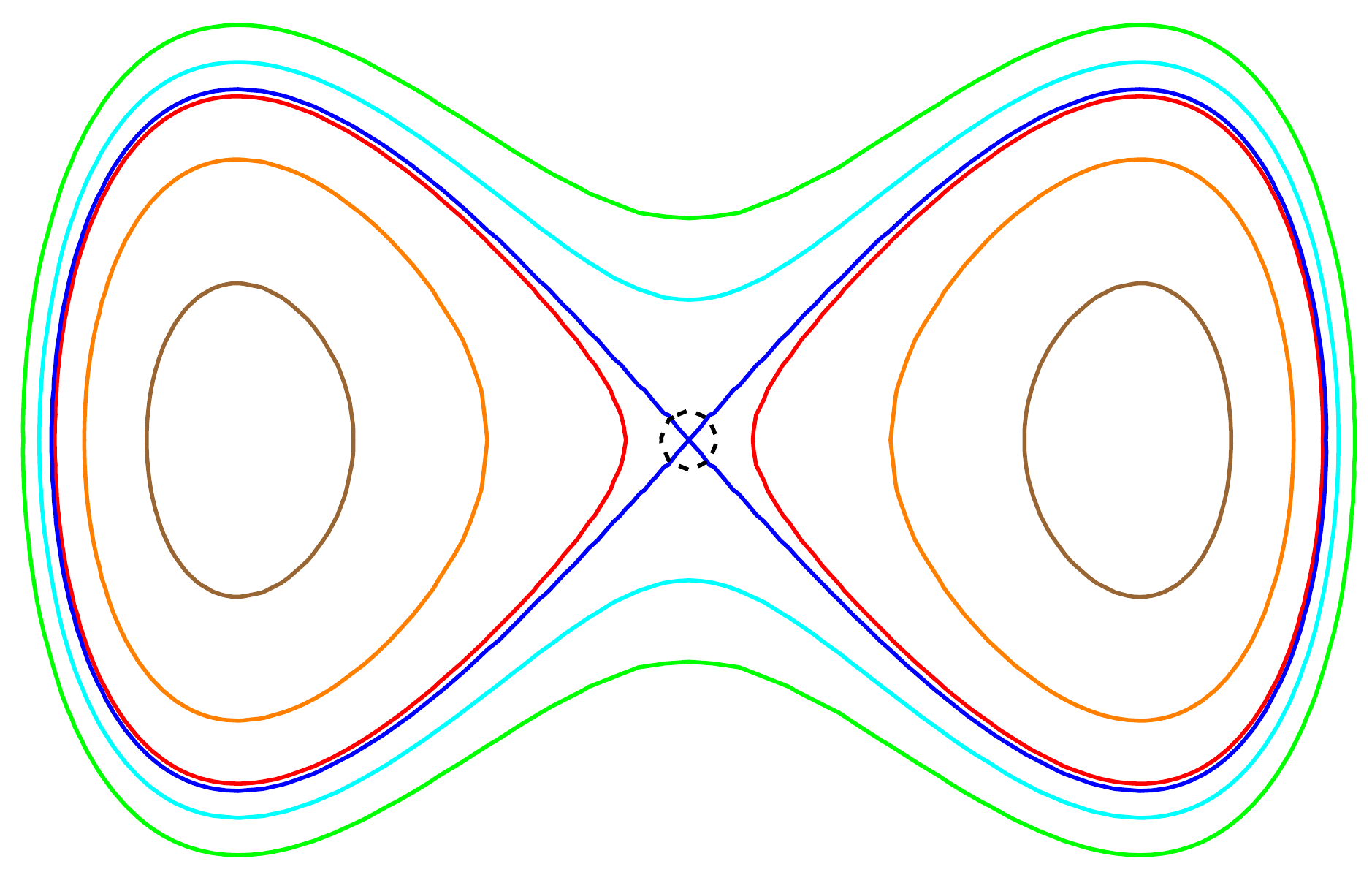}
        \caption{ELSs for the Ginzburg-Landau Hamiltonian. We picked six energy values within the range $\mathcal{E}:=[-0.089,1]$ such that each ELS is identified by a specific color. The blue curve represents $\Sigma_{E_c}^{H}$ with $E_c=0$ which admits a critical point identified by the small dashed circle. The brown, orange and red curves represent the ELSs $\Sigma_{E}^{H}$ for $E<E_c$ whereas the cyan and green curves those with $E>E_c$.}
    \label{Fig:energy_level_set_critical}
    \end{minipage}
\end{figure}

The second variation formula of volume can be obtained by differentiating once again Eq. \eqref{eqn:first_variation_volume} and we get:
\begin{equation}
    \begin{split}\label{eqn:2_variation_formula}
        \partial^{2}_{E}vol^{\widetilde{g}}(\Sigma^{H}_{E})&=\int_{\Sigma_{E}^{H}}\mathcal{L}_{\bm{\xi}}\bigg\{Tr[\mathcal{W}^{\widetilde{g}}_{\bm{\xi}}]\;d\eta^{\widetilde{g}}_{\Sigma^{H}_{E}}\bigg\}\\
        &=\int_{\Sigma_{E}^{H}}\bigg\{\partial_{E} Tr[\mathcal{W}^{\widetilde{g}}_{\bm{\xi}}]+Tr[\mathcal{W}^{\widetilde{g}}_{\bm{\xi}}]^{2}\bigg\}d\eta^{\widetilde{g}}_{\Sigma^{H}_{E}}.
    \end{split}
\end{equation}
Then, by exploiting both the trace of the Riccati equation \cite{zhou2013simple,gromov2019four}:
\begin{equation}\label{eqn:riccati_shape_operator}
    \partial_{E} Tr[\mathcal{W}^{\widetilde{g}}_{\bm{\xi}}]=-Tr[(\mathcal{W}^{\widetilde{g}}_{\bm{\xi}})^{2}]-Ric^{\widetilde{g}}(\bm{\xi},\bm{\xi}),
\end{equation}
where $Ric^{\widetilde{g}}(\bm{\xi},\bm{\xi})$ is the Ricci curvature tensor along the $\bm{\xi}$-direction and the Gauss-Codazzi equation \cite{gromov2019four}:
\begin{equation}\label{def:scalar_curv_relation}
\begin{split}
    Ric^{\widetilde{g}}(\bm{\xi},\bm{\xi})=\frac{1}{2}\bigg(R^{\widetilde{g}}(\Lambda)-R^{\widetilde{g}}(\Sigma_{E}^{H})
    +Tr[\mathcal{W}^{\widetilde{g}}_{\bm{\xi}}]^{2}-Tr[(\mathcal{W}^{\widetilde{g}}_{\bm{\xi}})^{2}]\bigg),
\end{split}
\end{equation}
Eq. \eqref{eqn:2_variation_formula} rewrites:
\begin{equation}\label{eqn:riccati_entropy_flow_geometric}
\begin{split}
    \partial^{2}_{E}vol^{\widetilde{g}}(\Sigma^{H}_{E})=\frac{1}{2}\int_{\Sigma_{E}^{H}}\bigg\{Tr[\mathcal{S}_{\bm{\xi}}]^{2}-Tr[\mathcal{S}^{2}_{\bm{\xi}}]
    +R^{g}(\Sigma_{E}^{H})-R^{g}(\Lambda)\bigg\}\;d\eta^{g}_{\Sigma_{E}^{H}}.
\end{split}
\end{equation}
It is therefore natural to define the following class of \emph{geometric curvature functions} (GCFs)
\begin{equation}\label{def:geometric_function}
    \Omega_{\widetilde{g}}^{(k)}(E):=\frac{\partial_{E}^{k}vol^{\widetilde{g}}(\Sigma_{E}^{H})}{vol^{\widetilde{g}}(\Sigma_{E}^{H})}.
\end{equation}
For $k=1$, we recover:
\begin{eqnarray}\label{def:first_order_geometric_func}
    \Omega_{\widetilde{g}}^{(1)}(E):=\int_{\Sigma_{E}^{H}}Tr^{\widetilde{g}}[\mathcal{W}^{\widetilde{g}}_{\bm{\xi}}]~d\rho^{\widetilde{g}}_{\Sigma_{E}^{H}},
\end{eqnarray}
whereas for $k=2$, we have:
\begin{equation}\label{def:second_order_geometric_func}
\begin{split}
    \Omega_{\widetilde{g}}^{(2)}(E):=\int_{\Sigma_{E}^{H}}\bigg\{Tr[\mathcal{W}^{\widetilde{g}}_{\bm{\xi}}]^{2}-Tr[(\mathcal{W}^{\widetilde{g}}_{\bm{\xi}})^{2}]
    +R^{\widetilde{g}}(\Sigma_{E}^{H})-R^{\widetilde{g}}(\Lambda)\bigg\}\;d\rho_{\Sigma_{E}^{H}}.
\end{split}
\end{equation}
By exploiting the notation introduced above, Eqs. \eqref{eqn:first_derivatives}, \eqref{eqn:second_derivatives} and \eqref{eqn:third_derivatives} reduce to 
\begin{equation}
        \partial_{E}S_{\widetilde{g}}(E)=\Omega_{\widetilde{g}}^{(1)}(E),\label{eqn:first_derivatives_g_gori}
\end{equation}
\begin{equation}
        \partial^{2}_{E}S_{\widetilde{g}}(E)=\Omega_{\widetilde{g}}^{(2)}(E)-(\Omega_{\widetilde{g}}^{(1)}(E))^{2},\label{eqn:second_derivatives_g_gori}
\end{equation}
\begin{equation}
\begin{split}
        \partial^{3}_{E}S_{\widetilde{g}}(E)=\Omega_{\widetilde{g}}^{(3)}(E)+2(\Omega_{\widetilde{g}}^{(1)}(E))^{3}-3\Omega_{\widetilde{g}}^{(2)}(E)\Omega_{\widetilde{g}}^{(1)}(E)\label{eqn:third_derivatives_g_gori}.
\end{split}
\end{equation}
In general, we have:
\begin{eqnarray}\label{def:entropy_omega}
        \partial_{E}^{k}S_{\widetilde{g}}(E)=\partial_{E}^{k-1}\Omega_{\widetilde{g}}^{(1)}(E).
\end{eqnarray}
The equations above are a key result since they manifest an exact equivalence between thermodynamics and geometry; therefore, a few remarks are in order. Equation \eqref{def:entropy_omega} is meaningful since \textit{(i)} our derivation is exact and thus we have a strong evidence that the only relevant mathematical structure at the basis of a PT has a geometric origin, contained in the scalar quantities $\Omega^{(k)}_{\widetilde{g}},\Omega^{(k-1)}_{\widetilde{g}},\ldots,\Omega^{(1)}_{\widetilde{g}}$. No further mathematical/physical entity plays a role in a PT. \textit{(ii)} The GCFs are physical observables that can be evaluated along the Hamiltonian dynamics exploiting the ergodic hypothesis. In fact, given any phase space-valued function, $f:\Lambda\rightarrow\mathbb{R}$, we have \cite{bel2021geometrical,gori2018topological}:
\begin{equation}
\begin{split}\label{def:time_average}
    \int_{\Sigma_{E}^{H}}f(\bm{x})\;d\rho^{\widetilde{g}}_{\Sigma_{E}^{H}}(\bm{x})\equiv \lim_{T\to\infty}\frac{1}{T}\int_{0}^{T} f(\bm{X}(\tau))\;d\tau,
\end{split}
\end{equation}
where $\bm{x}=\{\bm{q}^{1},\ldots,\bm{q}^{N},\bm{p}_{1},\ldots,\bm{p}_{N}\}$ whereas $\bm{X}(\tau)$ is the phase space-trajectory---solution of the Hamilton's equations of motion---which is computed numerically. \textit{(iii)} Any GCF corresponds to specific combinations of derivatives of the Hamiltonian functions (see Eqs. (S.9), (S.11) and (S.14) in the SI). Therefore, the information about a $k$ order PT is evidently enclosed into the geometric quantities $\{\Omega_{g}^{(n)}\}_{n=1}^{k}$, hence, a PT occurs depending on whether and how the ELS's geometry changes along the energy flow.  Equation \eqref{def:entropy_omega} allows us to reinterpret the Bachmann's classification from a purely geometric viewpoint. In fact, we can state that a PT of order $k$ à la Bachmann occurs at the transition energy point $E_t$ if the geometric observables $\partial_{E}^{k}\Omega_{\widetilde{g}}^{(1)}$ and $\partial_{E}^{k+1}\Omega_{\widetilde{g}}^{(1)}$, respectively, admit a least-sensitive inflection point and a positive-valued minimum (negative-valued maximum) in $E_t$. On the basis of Eq. \eqref{def:entropy_omega}, we now discuss the role of geometry for systems undergoing first- and second-order PTs à la Bachmann and, then, we investigate the behavior of the geometry in the thermodynamic limit discussing the connection between the Bachmann's and Ehrenfest's classifications. 
\paragraph{Role of the geometry} For a first-order PT, Eq. \eqref{eqn:first_derivatives_g_gori} shows that the entropy flow is guided by the average of $Tr[\mathcal{W}_{\bm{\xi}}^{\widetilde{g}}]$ which is the mean curvature (up to a constant). Therefore, the study of a first-order PT à la Bachmann is reduced to the evaluation of the total mean curvature along the Hamiltonian dynamics for each energy value. The so-obtained function is, then, plugged into Eq. \eqref{eqn:first_derivatives_g_gori} which, in turn, can be solved as a ordinary differential equation. This procedure has been employed for investigating the Ginzburg-Landau-like (GL) model (see section III in SI for further details). In practice, we numerically evaluated the total mean curvature \eqref{def:first_order_geometric_func} by means of Eq. \eqref{def:time_average} at each time step and we used it for integrating Eq. \eqref{eqn:first_derivatives_g_gori} numerically obtaining $S_{\widetilde{g}}(E)$. The results are plotted in Fig. \ref{Fig:GL_model}. The entropy function (orange curve) displays a least-sensitive inflection point in the entropy at the transition point $\epsilon_t=0$. In particular, the total mean curvature $\Omega_{\widetilde{g}}^{(1)}(E)$ has a positive-valued minimum on the left of $\epsilon_t$. Note that the apparent peak in $\Omega_{\widetilde{g}}^{(1)}(E)$ is due to the fact that the GL-model undergoes a PT in correspondence of a critical point which therefore enhances the back-bending region. In fact, this is the limiting case of the function $\beta(E)$ provided by Bachmann in Fig. \ref{Fig:bachmann} where the (local) maximum and the minimum points of $\beta$-function lie on the dashed vertical line. These results suggest that, in finite-size system, the origin of a first order PT can be traced back to a local growth, around the transition point, of the total mean curvature values associated to the ELSs as it can be deduced by the presence of a back-bending region in $\beta(E)(\equiv\partial_E S_{\widetilde{g}})$ (see Fig.~\ref{Fig:bachmann}). More precisely, this means that the ELSs around the transition energy point contain subsets of non-vanishing measure where the mean curvature takes large values. This occurs, for example, when a hypersurface develops cusp-shaped subsets as in the GL-model (see Fig.~\ref{Fig:energy_level_set_critical}). We will discuss this mechanism later on.
For a second-order PT, we exploit Eq. \eqref{eqn:first_derivatives_g_gori} in Eq. \eqref{eqn:second_derivatives_g_gori} so as to obtain:
\begin{equation}\label{eqn:riccati_entropty}
    \partial^2_E S_{\widetilde{g}}(E)+(\partial_E S_{\widetilde{g}}(E))^2=\Omega_{\widetilde{g}}^{(2)}(E).
\end{equation}
This can be interpreted as a Riccati differential equation where the entropy is the unknown function. Interestingly, since the first-order GCF does not enter Eq. \eqref{eqn:riccati_entropty}, this suggests that all the information about a second-order PT is fully contained in the average of $Tr[\mathcal{W}^{\widetilde{g}}_{\bm{\xi}}]^{2}-Tr[(\mathcal{W}^{\widetilde{g}}_{\bm{\xi}})^{2}]+R^{\widetilde{g}}(\Sigma_{E}^{H})-R^{\widetilde{g}}(\Lambda)$. Therefore, we can study a second-order PT employing the same procedure developed for the first-order one. Thus, we reduced a thermodynamic problem to a real-analysis one. We evaluate $\Omega_{\widetilde{g}}^{(2)}$, regarded as a function of energy, and integrate Eq. \eqref{eqn:riccati_entropty} so as to obtain the functional dependence of the entropy function and its derivatives by the energy. Finally, these behaviors can be compared with those predicted by Bachmann in Fig. \ref{Fig:bachmann}. Such procedure will be implemented in a future work. In this Letter, instead, we just show the energy behaviors of the first- and second-order GCFs in the $\phi^4$-model (see SI for further details about the simulations). Essentially, we evaluated both $\Omega_{\widetilde{g}}^{(1)}$ and $\Omega_{\widetilde{g}}^{(2)}$ exploiting Eq. \eqref{def:time_average} as in the GL model and they are reported, respectively, in Figs.~\ref{Fig:phi4_model} and \ref{Fig:phi4_omega2}.
As a proof of concept, we inferred the second-order derivative of the entropy composing the GCFs as suggested by the right-hand side of Eq. \eqref{eqn:second_derivatives_g_gori}. This results is reported in Fig. \ref{Fig:phi4_model}. The details about the numerical evaluation of the first- and second-order GCFs can be found in the SI. We observe a least-sensitive inflection point both in $\Omega_{\widetilde{g}}^{(1)}$ and $\Omega_{\widetilde{g}}^{(2)}$ at the transition point $\epsilon_t=11.1$ and a negative-valued peak in $\partial_{E}\Omega_{\widetilde{g}}^{(1)}$ still in $\epsilon_t=11.1$. Note that the energy-behaviors of $\Omega_{\widetilde{g}}^{(1)}$ and $\Omega_{\widetilde{g}}^{(2)}-(\Omega_{\widetilde{g}}^{(1)})^2$ are the same as those predicted by Bachmann (see the functions $\beta(E)$ and $\gamma(E)$ in Fig. \ref{Fig:bachmann}). Therefore, we conclude that \textit{(i)} the finite-size $\phi^4$-model undergoes a second-order PT à la Bachmann around $\epsilon_t\approx 11.1$, \textit{(ii)} this PT is triggered by a change of geometry as can be deduced from the GCFs' behaviors. It is worth noting that our results are fully in agreement with those obtained in Ref. \cite{bel2020geometrical}.
Surprisingly, Eqs. \eqref{eqn:first_derivatives_g_gori} and \eqref{eqn:riccati_entropty} suggest a sort of universality in the energy behaviors of the first- and second-order GCFs, at least, in a neighborhood of the transition energy point. In fact, it can be proved that (see SI, section IV) all the Hamiltonian systems which undergo a first- (second-) order PT à la Bachmann manifest the same qualitative behavior in the first- (second-) order GCF. More specifically, let us suppose to have a well-precise behavior in $\Omega_{\widetilde{g}}^{(1)}$ ($\Omega_{\widetilde{g}}^{(2)}$) so as to produce the correct behavior of the entropy as predicted by Bachmann. By exploiting \textit{(i)} the Cauchy’s theorem of existence and uniqueness, \textit{(ii)} the constraint given by the Bachmann's classification, one can show that if there exists another first- (second-) order GCF such that to satisfy Eq. \eqref{eqn:first_derivatives_g_gori} (Eq. \eqref{eqn:riccati_entropty}), then, this is qualitatively equal to the previous one. Finally, it is worth analyzing what is the role of the critical points in this context. As discussed in section \textit{a.}, the mere presence of a critical point does not ensure the discontinuity of entropy and, thereby, the occurrence of a PT. In general, we expect that, in order for $\bm{x}_c$ to give rise to a PT, it must deeply affect the geometry of the respective ELS. For instance, let us analyze what happens in case of a first-order PT. Let us assume that to the energy value $E_c\in\mathcal{E}$ is associated an ELS, $\Sigma_{E_c}^{H}$, containing a critical point, $\bm{x}_c$, and let us consider the range of energies $\mathcal{A}_{\epsilon}:=[E_c-\epsilon,E_c+\epsilon]$ for a fixed $\epsilon>0$. Then, let us pick a subset $\mathcal{U}\subset\Sigma_{E_c-\epsilon}^{H}$ such that, under the flow $E_c-\epsilon\to E_c$, it is transformed into a neighborhood, $\mathcal{U}_{\bm{x}_c}\subset\Sigma_{E_c}^{H}$, of $\bm{x}_c$ of non-vanishing measure, i.e.:
\begin{equation}
    vol^{\widetilde{g}}(\mathcal{U}_{\bm{x}_c})=\int_{\mathcal{U}_{\bm{x}_c}}d\eta^{\widetilde{g}}_{\Sigma_{E_c}^{H}}\neq 0.
\end{equation}

In so doing, we have a collection of sets $\{\mathcal{U}_{E}\}_{E\in\mathcal{A}_{\epsilon}}$ such that for $E=E_c$, we get $\mathcal{U}_{E_c}\equiv\mathcal{U}_{\bm{x}_c}$.
Then, noting that $\beta(E)\equiv\partial_E S_{\widetilde{g}}=\Omega_{\widetilde{g}}^{(1)}$, the emergence of a back-bending region in $\beta(E)$ can be due to the presence of a critical point such that, along the evolution $\Sigma_{E_c-\epsilon}\to\Sigma_{E_c}^{H}$, it joints two parts of an ELS into a single one, $\Sigma_{E_c}^{H}$ (see in Fig.~\ref{Fig:energy_level_set_critical}, the red curves which merge into the blue one). During this hypersurfaces-merging process, the critical point coincides with the connection point of the two merged parts of $\Sigma_{E_c}^{H}$ (see the dashed circle in Fig.~\ref{Fig:energy_level_set_critical}). Hence, below the energy threshold $E_c$, the contribution to the total mean curvature values given by the subsets, $\{\mathcal{U}_E\}_{E<E_c}$, is larger and larger until reaching its largest value on $\mathcal{U}_{\bm{x}_c}$. Above the energy threshold, instead, the process is reversed, that is, the contribution given by $\{\mathcal{U}_E\}_{E>E_c}$ is smaller and smaller and in the $E\to\infty$ limit, the total mean curvature vanishes. We emphasize that this is not the only possible process which gives rise to a first-order PT à la Bachmann. Moreover, the presence of a critical point seems to be neither necessary nor sufficient as it has been shown by Kastner and co-workers in Refs. \cite{mehta2012energy,kastner2011phase}. In fact, in the $\phi^4$-model, the critical points can be found at energies remarkably far from the transition one, thereby, they do not contribute at all to the emergence of a PT. In conclusion, accordingly with the previous discussion, the absence of a PT in correspondence of critical points can be due to the fact that the neighborhoods, where the mean curvature (or, eventually, the scalar curvature) should contribute with large values, have actually zero-measure.\\
\paragraph{Geometry's behavior in the thermodynamics limit.} Finally, by exploiting an analysis based on the differentiability class of the microcanonical entropy, we show that Eqs. \eqref{eqn:first_derivatives_g_gori} and \eqref{eqn:riccati_entropty} provide an insight about the geometric origin of first- and second-order PTs for systems that admit a thermodynamic limit. To this purpose, we adopt the Ehrenfest-like classification developed in Ref. \cite{bel2020geometrical} for the microcanonical ensemble and we make explicit the dependence of any observable by the number of DoF, $n$. Essentially, it is well-known that the canonical free energy, $f_{n}(\beta)$, with $\beta=1/k_{B}T$, is related to the microcanonical entropy through the Legendre transform:
\begin{equation}\label{def:legendre_free_energy}
    -f_{n}(\beta)=\beta E-S_{\widetilde{g},n}(E),
\end{equation}
Then, let us denote with $\beta_{n}(E):=\partial S_{\widetilde{g},n}(E)/\partial E$ the inverse of the canonical temperature $T(E)$ and use such a relation in the argument of the free energy in Eq. \eqref{def:legendre_free_energy} so that $F_{n}(E):=f_{n}(\beta_{n}(E))$. It is easy to show that $\partial F_{n}(E)/\partial E=-E\,\partial\beta_{n}(E)/\partial E$, thereby, if $\beta_{n}\in C^{k}(\mathcal{E},\mathbb{R})$ then $F_{n}\in C^{k}(\mathcal{E},\mathbb{R})$ implying that $S_{\widetilde{g},n}\in C^{k+1}(\mathcal{E},\mathbb{R})$. By assuming the existence of the Legendre transform in the $n\to\infty$ limit (we refer to Ref. \cite{bel2020geometrical} for a deeper discussion), we associate a first- or second-order PT à la Ehrenfest in the microcanonical ensemble if the second- or third-derivative of entropy $S_{\widetilde{g},n}$ admits a discontinuity for a certain energy value, $E_t$. 
In a first-order PT, the sequence of functions $\{\partial_E S_{n}\}_{n\in\mathbb{N}}$ uniformly converges to a function $\partial_E S_{\infty}\in C^{0}(\mathcal{E},\mathbb{R})$ such that $\partial^2_E S_{\infty}$ is discontinuous in $E_t\in\mathcal{E}$; then, by inspection of Eq. \eqref{eqn:first_derivatives_g_gori}, we deduce that $\{\Omega_{\widetilde{g},n}^{(1)}\}_{n\in\mathbb{N}}$ does uniformly converge to $\Omega_{\widetilde{g},\infty}^{(1)}\in C^{0}(\mathcal{E},\mathbb{R})$ and exploiting Eq. \eqref{eqn:riccati_entropty} we conclude that $\Omega_{\widetilde{g},\infty}^{(2)}$ must be discontinuous in $E_t$. We expect that such a discontinuity arises from $R^{\widetilde{g}}(\Sigma_{E}^{H})$ but we leave the rigorous proof of such a guess to future works. \\
For a second-order PT, we have that the sequences of functions $\{\partial_{E}S_n\}_{n\in\mathbb{N}}$ and $\{\partial_{E}^{2}S_n\}_{n\in\mathbb{N}}$ uniformly converge, respectively, to $\partial_{E}S_\infty\in C^1(\mathcal{E},\mathbb{R})$ and $\partial_{E}^{2}S_\infty\in C^0(\mathbb{R})$. Then, from Eq. \eqref{eqn:riccati_entropty}, we deduce that the sequence $\{\Omega^{(2)}_{\widetilde{g},n}\}_{n\in\mathbb{N}}$ must converge uniformly to $\Omega^{(2)}_{\widetilde{g},\infty}\in C^0(\mathcal{E},\mathbb{R})$, since $\partial_{E}S_\infty(E)\in C^1(\mathcal{E},\mathbb{R})$. Finally, by exploiting Eq. \eqref{eqn:third_derivatives_g_gori}, we conclude that $\Omega^{(3)}_{\widetilde{g},\infty}(E)$ must converge to a function with a discontinuity in $E_t$. In this case, we may expect that such a discontinuity stems from $\partial_E R(\Sigma_{E}^{H})$. \\
It should be noted that discontinuities in the entropy derivatives, at finite $n$, have been observed in many systems such as, for example, two particles interacting through the Lennard-Jones potential \cite{hilbert2006nonanalytic,dunkel2006phase}, in the p-trig model \cite{angelani2005topology,pettini2007geometry}, in the mean-field-like Berlin-Kac model \cite{kastner2006mean,casetti2006nonanalyticities} and in the $\phi^4$-model on the real line \cite{kastner2008phase}. Such non-analyticities that, in some case, are due to critical points and so to topology changes, cannot be conceptually related to the occurrence of PTs in the standard sense of the statistical mechanics. In fact, the number of these discontinuities may grow exponentially with the number of DoF, $n$, meaning that the first $k$ derivatives of the entropy are continuous, where $k$ is of order $n$. Moreover, their ``strength'' generically decreases linearly with $n$, thereby, in the thermodynamic limit most of these non-analyticities can disappear \cite{kastner2008nonanalyticities,kastner2008phasevanishing,casetti2009kinetic,nerattini2013exploring}.

%%%%%%%%%%%%%%%%%%%%%%%%%%%%%%%

\paragraph*{Conclusion.}
In this Letter, we proposed a geometric theory of PTs that provides a rich and coherent description of the transitional phenomena in the microcanonical ensemble. By rephrasing the microcanonical entropy as the logarithm of the ELS's volume, we derived a class of exact relations which connect the derivatives of the microcanonical entropy to the geometric curvature observables of the ELSs. More specifically, we showed that any entropy derivative of order $k$ is actually affected by a well-precise geometric entities, for instance, the energy variation of $\partial_E S_{\widetilde{g}}$ is due to a variation of the total mean curvature and similarly for $\partial^2_E S_{\widetilde{g}}$. Therefore, such an approach allows us to interpret the emergence of a PT as a peculiar change of a specific geometric property of the ELSs. For physical systems which admit the thermodynamic limit, we have shown that, in order to have a Ehrenfest-like PT of order $k$, a loss of analyticity must manifest at the level of the $k$-order GCF and this corresponds to a specific geometry change. However, non-analyticities of the ELSs' geometry do not necessarily show up at low numbers of DoF, thereby, any Ehrenfest-like classification is unappropriated. Nevertheless, by adopting the Bachmann's classification, it is still possible \textit{(i)} to determine whether a PT is ongoing, \textit{(ii)} to recognize a geometry change of the ELSs that, however, is weaker than that expected in the infinite-size system. It should be stressed that, although the geometric observables such as the scalar curvature and the mean curvature seem to be just formal mathematical structures, they are actually related to specific combination of derivatives of the Hamiltonian function. Therefore, we conclude that this approach is of general validity and it can be employed for investigating PTs, in particular of first- and second-order, that occur in any autonomous Hamiltonian system independently by the number of DoF.
%From a geometric viewpoint, just knowing the geometric evolution of the ELSs, we can infer whether or not thesystem undergoes a PT.\\

% ===== Plots =====

% Energy level sets GL

\begin{figure}[!htb]
    
\end{figure}

\paragraph*{Acknowledgments.}
I am indebted with Dr. Matteo Gori and Prof. Marco Pettini for their fundamental advises without which I would not have reached these results. I also thank Dr. Riccardo Capelli for precious suggestions about numerics and for a careful reading of this work. 
%I wish to express my gratitude to the anonymous referees whose suggestions and criticisms have raised the quality of this Letter. 
Finally, this work is dedicated to the memory of my mentor, Riccardo Ignazzi.

% =========== SUPPLEMENTARY INFORMATION ============

\pagebreak

\textbf{SUPPLEMENTARY INFORMATION:\\
The Geometric Theory of Phase Transitions}

\author{Loris Di Cairano}

\address{Department of Physics and Materials Science, University of Luxembourg, L-1511 Luxembourg City, Luxembourg and}
\address{Computational Biomedicine, Institute of Neuroscience and Medicine INM-9 and Institute for Advanced Simulations IAS-5, Forschungszentrum Jülich, 52428 Jülich, Germany.}

 \ead{l.di.cairano.92@gmail.com, loris.dicairano@uni.lu}

\setcounter{equation}{0}

\section{Computation of the geometric observables}

We consider a generic autonomous Hamiltonian system, described by the Hamiltonian function $H:\Lambda\subset\mathbb{R}^{N}\times\mathbb{R}^{N}\rightarrow\mathbb{R}$ where $\Lambda$ is the phase space. It is well-known that, picking a fixed energy value $H(\bm{x})=E$, the dynamics of the representative point of the system, $\bm{x}=\{\bm{p},\bm{q}\}\in\Lambda$, lies on the energy level set \cite{pettini2007geometry} ($n=2N$):
\begin{equation}\label{def:energy_level_sets}
    \Sigma^{H}_{E}:=\{\bm{x}\in \Lambda\,|\, H(\bm{x})=E\}\subset\mathbb{R}^n.
\end{equation} 
By allowing the energy to vary over a set of accessible energies denoted by $\mathcal{E}:=[E_0,E_1]$, we obtain a collection $\{\Sigma_{E}^{H}\}_{E\in\mathcal{E}}$ of energy level sets which defines a foliation of the phase space, i.e., $\Lambda=\bigcup_{E\in\mathcal{E}}\;\Sigma_{E}^{H}$.
We thus need to introduce an appropriate metric tensor on the phase space which, in turn, induces a metric tensor, $h_{E}$, on each hypersurface. Now, from a purely mathematical viewpoint, we have no restriction about the specific structure of the metric tensor and this led to introduce in Ref. \cite{di2021topology} the metric tensor
\begin{equation}\label{def:metric_tensor_gromov}
    g = dE\otimes dE+h_E,
\end{equation}
where $E$ is interpreted as an energy parameter and $h_E$ is the Euclidean metric tensor induced on the hypersurface $\Sigma_{E}^{H}$.\\
By adopting the metric tensor \eqref{def:metric_tensor_gromov}, we can study the extrinsic geometry of each energy level set through the introduction of the Weingarten operator
\begin{equation}\label{def:weingarten_g}
    \mathcal{W}^{g}_{\bm{\nu}}(\bm{X}):=\nabla_{\bm{X}}\bm{\nu}\equiv\nabla\left(\frac{\nabla H}{\|\nabla H\|}\right)\cdot\bm{X},
\end{equation}
where $\bm{X}$ is any vector field lying on the tangent space to the energy level set $\Sigma_{E}^{H}$, $\nabla$ and $\|\cdot\|$ are, respectively, the gradient operator and the norm defined on the phase space.
The phase space can be equipped with a more appropriate metric tensor \cite{gori_thesis,gori2018topological} defined by
\begin{equation}\label{def:conformal_metric}
    \widetilde{g}=dE\otimes dE+\chi^{\frac{2}{n-1}}(h_E)_{ij}~dx^{i}\otimes dx^{j},
\end{equation}
where $\chi:=1/\|\nabla H\|$. The collection of energy level sets can be generated evolving the energy level set $\Sigma_{E_0}^{H}$ through the vector field $\partial_{E}$ such that $dE(\partial_{E})=1$ \cite{hirsch2012differential}. In particular, it is easy to show that \cite{di2021topology}:
\begin{equation}\label{def:zeta_vector}
    \bm{\xi}:=\partial_{E}=\chi\bm{\nu}.
\end{equation}
We note that the vector field $\bm{\xi}$ has the following properties: \textit{(i)} it generates the flow so that $dE(\partial_{E})=1$, \textit{(ii)} it is normalized to the unity in the metric tensor \eqref{def:conformal_metric}. This allows us to introduce a new Weingarten operator defined by
\begin{equation}\label{def:weingarten_g_tilde}
    \mathcal{W}^{\widetilde{g}}_{\bm{\xi}}(\bm{X}):=\nabla^{\widetilde{g}}_{\bm{X}}\bm{\xi}\cdot \bm{X}\equiv\nabla\left(\frac{\nabla H}{\|\nabla H\|^{2}}\right)\cdot \bm{X}
\end{equation}
where $\nabla^{\widetilde{g}}$ is defined with respect to the metric tensor $\widetilde{g}$ whereas $\nabla$ is the gradient operator on the phase space as defined before.\\
The Weingarten operator \eqref{def:weingarten_g_tilde} is related to the previous one (see Eq. \eqref{def:weingarten_g}) through the relation
\begin{equation}\label{def:transf_wing}
    \mathcal{W}^{g}_{\bm{\nu}}\mapsto   \mathcal{W}^{\widetilde{g}}_{\bm{\xi}}=\chi~\mathcal{W}^{g}_{\bm{\nu}}+\chi^{-1}\partial_{E}\chi\frac{1\!\!\!1_{\Sigma^{H}_{E}}}{n-1}, 
\end{equation}
where 
\begin{equation}
    \partial_{E}\chi=\langle \bm{\xi},\nabla\chi\rangle_{\widetilde{g}}=\chi\langle \bm{\nu},\nabla\chi\rangle_{\widetilde{g}}.
\end{equation}
and $1\!\!\!1_{\Sigma_{E}^{H}}$ is the identity operator defined on the tangent space to $\Sigma_{E}^{H}$.

\subsection{Expression of the geometric observables}

We now derive the explicit expression for the geometric curvature functions (GCFs) adopting the metric tensor $\widetilde{g}$ together with the expression \eqref{def:transf_wing} for the Weingarten.\\
We recall below the expression for the first and second order GCFs as reported in the main text:  
\begin{equation}
\begin{split}\label{def:Omega_1_2}
    \Omega_{\widetilde{g}}^{(1)}(E)&:=\int_{\Sigma_{E}^{H}}Tr[\mathcal{W}^{\widetilde{g}}_{\bm{\xi}}]~d\rho\\
    \Omega^{(2)}_{\widetilde{g}}(E)&:=\frac{1}{2}\int_{\Sigma_{E}^{H}}\bigg\{Tr[\mathcal{W}^{\widetilde{g}}_{\bm{\xi}}]^{2}-Tr[(\mathcal{W}^{\widetilde{g}}_{\bm{\xi}})^{2}]
    +R^{\widetilde{g}}(\Sigma_{E}^{H})-R^{\widetilde{g}}(\Lambda)\bigg\}\;d\rho,
\end{split}
\end{equation}
Thus, the Weingarten operator is
\begin{equation}
    \mathcal{W}^{\widetilde{g}}_{\bm{\xi}}=\frac{Hess\, H}{\|\nabla H\|^{2}}-2\frac{\nabla H\otimes Hess\, H\cdot\nabla H}{\|\nabla H\|^{4}},
\end{equation}
so that the trace reads:
\begin{equation}
    Tr[\mathcal{W}^{\widetilde{g}}_{\bm{\xi}}]=\frac{\Delta H}{\|\nabla H\|^{2}}-2\frac{\langle\nabla H, Hess\, H\cdot\nabla H\rangle}{\|\nabla H\|^{4}}.
\end{equation}
For what concerns the second order GCF, this can be explicitly computed from its definition:
\begin{equation}
    \Omega_{\widetilde{g}}^{(2)}(E):=\frac{\partial^2_{E}vol^{\widetilde{g}}(E)}{vol^{\widetilde{g}}(E)}.
\end{equation}
In other words, we compute
\begin{equation}
    \partial^2_{E}vol^{\widetilde{g}}(E)=\int_{\Sigma_{E}^{H}}\partial_{E}\left(Tr[\mathcal{W}^{\widetilde{g}}_{\bm{\xi}}]d\mu^{\widetilde{g}}\right),
\end{equation}
which yields
\begin{equation}
\begin{split}\label{def:omega_function}
    \Omega^{(2)}_{\widetilde{g}}(E):=\int_{\Sigma_{E}}\bigg\{\frac{Tr[\mathcal{W}^{g}_{\bm{\nu}}]^2-Tr[(\mathcal{W}^{g}_{\bm{\nu}})^2]}{\|\nabla H\|^{2}}&-3\frac{Tr[\mathcal{W}^{g}_{\bm{\nu}}]}{\|\nabla H\|^{5}}\langle\nabla H,HessH\cdot\nabla H\rangle_{g^{\mathbb{E}}}-\frac{\nabla\nabla\nabla H(\nabla H,\nabla H,\nabla H)}{\|\nabla H\|^{6}}\\
    &+\frac{5}{\|\nabla H\|^{8}}\langle\nabla H,HessH\cdot\nabla H\rangle^{2}_{g^{\mathbb{E}}}-2\frac{\|HessH\cdot\nabla H\|^{2}}{\|\nabla H\|^{6}} \bigg\}d\rho^{\widetilde{g}},
\end{split}
\end{equation}
where $\mathcal{W}_{\bm{\nu}}^{g}$ is the Weingarten operator defined in Eq. \eqref{def:weingarten_g} whose traces are
\begin{equation}
\begin{split}
    Tr[\mathcal{W}^{g}_{\bm{\nu}}]&=\frac{\Delta H}{\|\nabla H\|}-\frac{\langle\nabla H, Hess\, H\cdot\nabla H\rangle}{\|\nabla H\|^{3}},\\
    Tr[(\mathcal{W}^{g}_{\bm{\nu}})^2]&=\frac{Tr[Hess\,H^2]}{\|\nabla H\|^2}+\frac{\langle \nabla H,Hess\,H\cdot\nabla H\rangle^2}{\|\nabla H\|^6}-2\frac{\|Hess\,H\cdot\nabla H\|^2}{\|\nabla H\|^4}.
\end{split}
\end{equation}
Note that the trace of the square of the Weingarten operator $\mathcal{W}^{\widetilde{g}}_{\bm{\xi}}$ is
\begin{equation}
    Tr[(\mathcal{W}^{\widetilde{g}}_{\bm{\xi}})^2]=\frac{Tr[Hess\;H^2]}{\|\nabla H\|^4}
    + 4 \frac{\langle \nabla H,Hess\;H\nabla H\rangle^2}{\|\nabla H\|^8} - 4\frac{\|Hess\;H\nabla H\|^2}{\|\nabla H\|^6}.
\end{equation}

%%%%%%%%%%%%%%%%%%%%%%%%%

\section{Numerical details for the $\phi^{4}$-model}

%%%%%%%%%%%%%%%%%%%

\subsection{Integration of the Hamilton's equation and computation of the thermodynamic observables}
\label{app:thermodynamics}

We performed molecular dynamics simulations in the microcanonical ensemble for the 2-dimensional $\phi^{4}$ Hamiltonian system with $N=30^{2}=900$ particles. The Hamiltonian is
\begin{equation}
\begin{split}\label{def:hamiltonian_phi4}
    H:=\sum_{\bm{i}}\Bigg[\frac{\pi^{2}_{\bm{i}}}{2}+\frac{\lambda}{4!}\phi^{4}_{\bm{i}}&-\frac{\mu^{2}}{2}\phi^{2}_{\bm{i}}+\frac{J}{4}\sum_{\bm{k}\in N\!\!N(\bm{i})}(\phi_{\bm{i}}-\phi_{\bm{i}})^{2}\Bigg]
\end{split}
\end{equation}
with $\lambda=3/5$, $\mu=2$ and $J=1$, then, $\bm{i}:=(i_1,i_2)$ is the two-dimensional index for labeling the sites, i.e., $1\leq i_1\leq n=30$ and $1\leq i_2\leq n=30$; finally $N\!\!N(\bm{i})$ is the set of the nearest neighbour lattice sites associated to $\bm{i}$.\\
The numerical integration of the Hamilton's equations of motion derived from the Hamiltonian function \eqref{def:hamiltonian_phi4} has been performed, with periodic boundary conditions, choosing random initial conditions and using a bilateral symplectic integration scheme \cite{casetti1995efficient,gori2018topological} with a time step such that the energy is conserved within the relative precision of $\Delta E/E\approx 10^{-6}$. The $2D$-$\phi^{4}$ model undergoes a phase transition at the critical energy density value $\epsilon_{t}=E_{t}/N\approx 11.1$ as already observed in Refs. \cite{bel2020geometrical,gori2018topological}. Note that by working with the potential function, instead of the Hamiltonian one, the critical potential density value is $v_{t}\approx 2.2$ as found in Refs. \cite{mehta2012energy,kastner2011phase,caiani1998hamiltonian}. Finally, since the study of the thermodynamic observables' behaviors as function of the number of the degrees of freedom, $N$, has been already done in Ref. \cite{bel2020geometrical}, we restricted our analysis to a system with $900$ particles. We evaluated the geometric observables in Eq. \eqref{def:Omega_1_2} exploiting the ergodic hypothesis, that is, we converted the microcanonical averages into time averages \cite{bel2021geometrical,gori2018topological}. More precisely, given a generic phase-space-valued function, $f:\Lambda\rightarrow\mathbb{R}$, we have
\begin{equation}
\begin{split}\label{def:time_average}
    \langle f(\Sigma_{E}^{H})\rangle_{M}:=\int_{\Sigma_{E}^{H}}f\;d\rho\equiv \lim_{T\to\infty}\frac{1}{T}\int_{0}^{T} f(\bm{X}(\tau))\;d\tau=:\overline{f(\Sigma_{E}^{H})},
\end{split}
\end{equation}
where we recall that $\bm{x}=\{\bm{q}^{1},\ldots,\bm{q}^{N},\bm{p}_{1},\ldots,\bm{p}_{N}\}$ whereas $\bm{X}(\tau)$ is the phase space-trajectory---solution of the Hamilton's equations of motion---which has been computed numerically.\\
By means of Eq. \eqref{def:time_average}, we evaluated $\Omega_{\widetilde{g}}^{(1)}$ and $\Omega_{\widetilde{g}}^{(2)}$ along the Hamiltonian dynamics for each energy density value within the set $[0.5,40]$. The results are plotted in the main text. We observe a least sensitive inflection point in $\Omega_{\widetilde{g}}^{(1)}$ at the transition point $\epsilon_t=11.1$ and a negative-valued peak in $\partial_{E}^2 S_{\widetilde{g}}=\Omega_{\widetilde{g}}^{(2)}-(\Omega_{\widetilde{g}}^{(1)})^2$ still in $\epsilon_t=11.1$. Note that the energy-behaviors of both $\partial_{E} S_{\widetilde{g}}=\Omega_{\widetilde{g}}^{(1)}$ and $\partial_{E}^2 S_{\widetilde{g}}$ are qualitatively the same as those of $\beta(E)$ and $\gamma(E)$ in Fig. 4 (main text) predicted by Bachmann. Finally, we note that these results are in agreement with those obtained in Ref. \cite{bel2020geometrical}.

\begin{figure}[!htb]
    \begin{minipage}[t]{.3\textwidth}
        \centering
        \includegraphics[width=\textwidth]{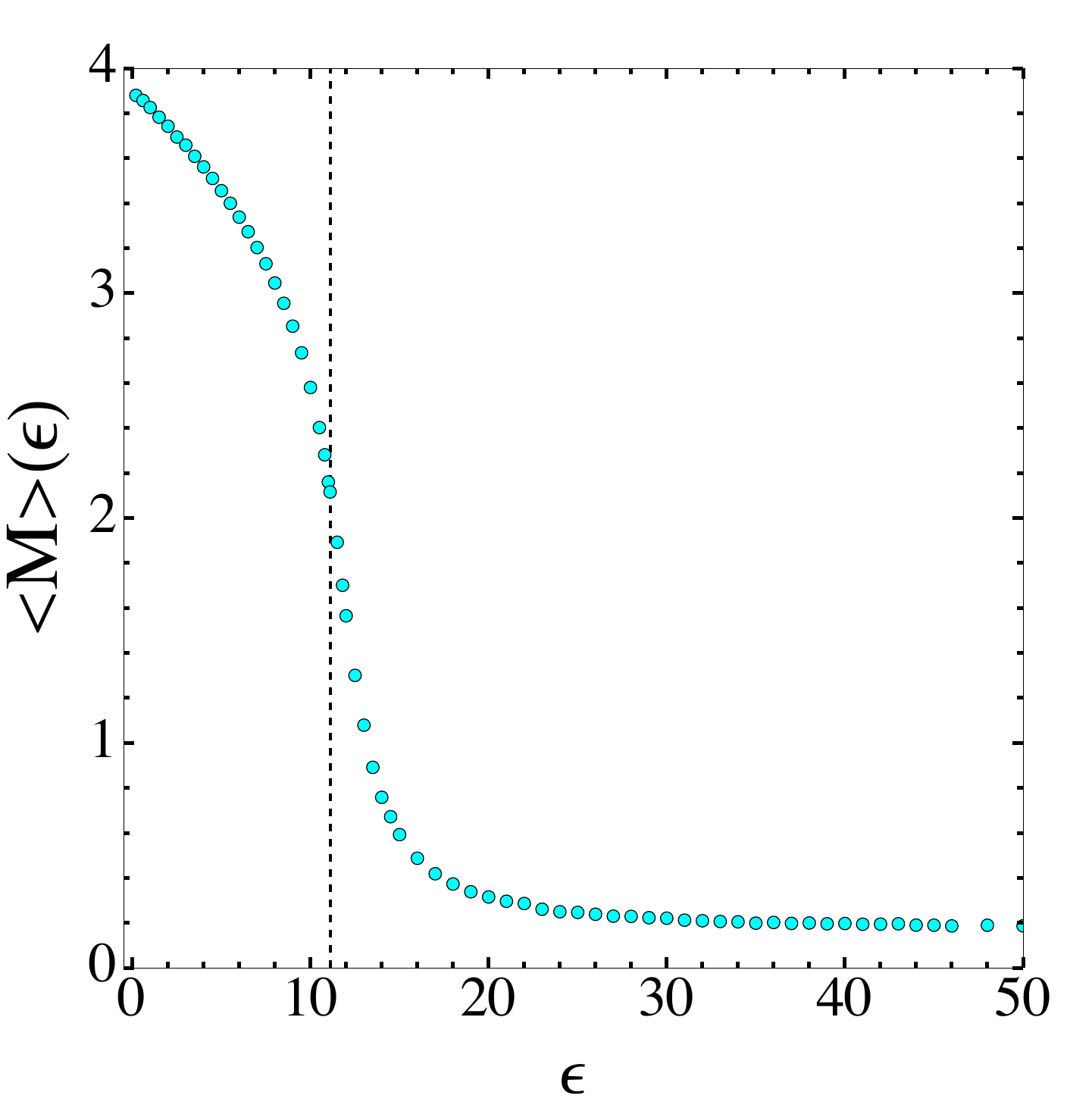}
       \caption{Magnetization.}
        \label{SFig:magnetization}
    \end{minipage}
        \begin{minipage}[t]{.3\textwidth}
        \includegraphics[width=\textwidth]{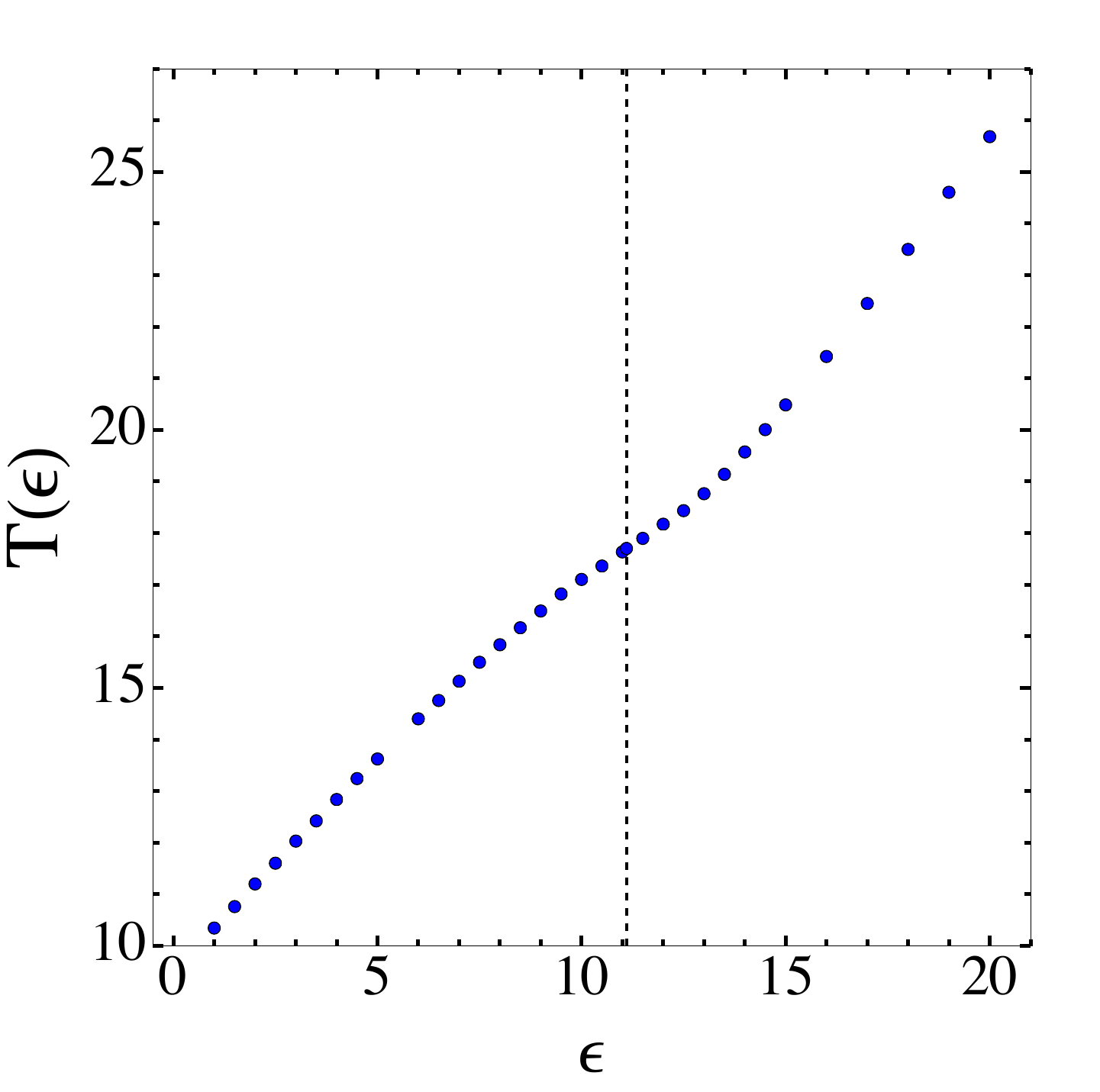}
        \caption{Caloric curve. }
        \label{SFig:caloric_curve}
    \end{minipage}
    \begin{minipage}[t]{.3\textwidth}
        \includegraphics[width=\textwidth]{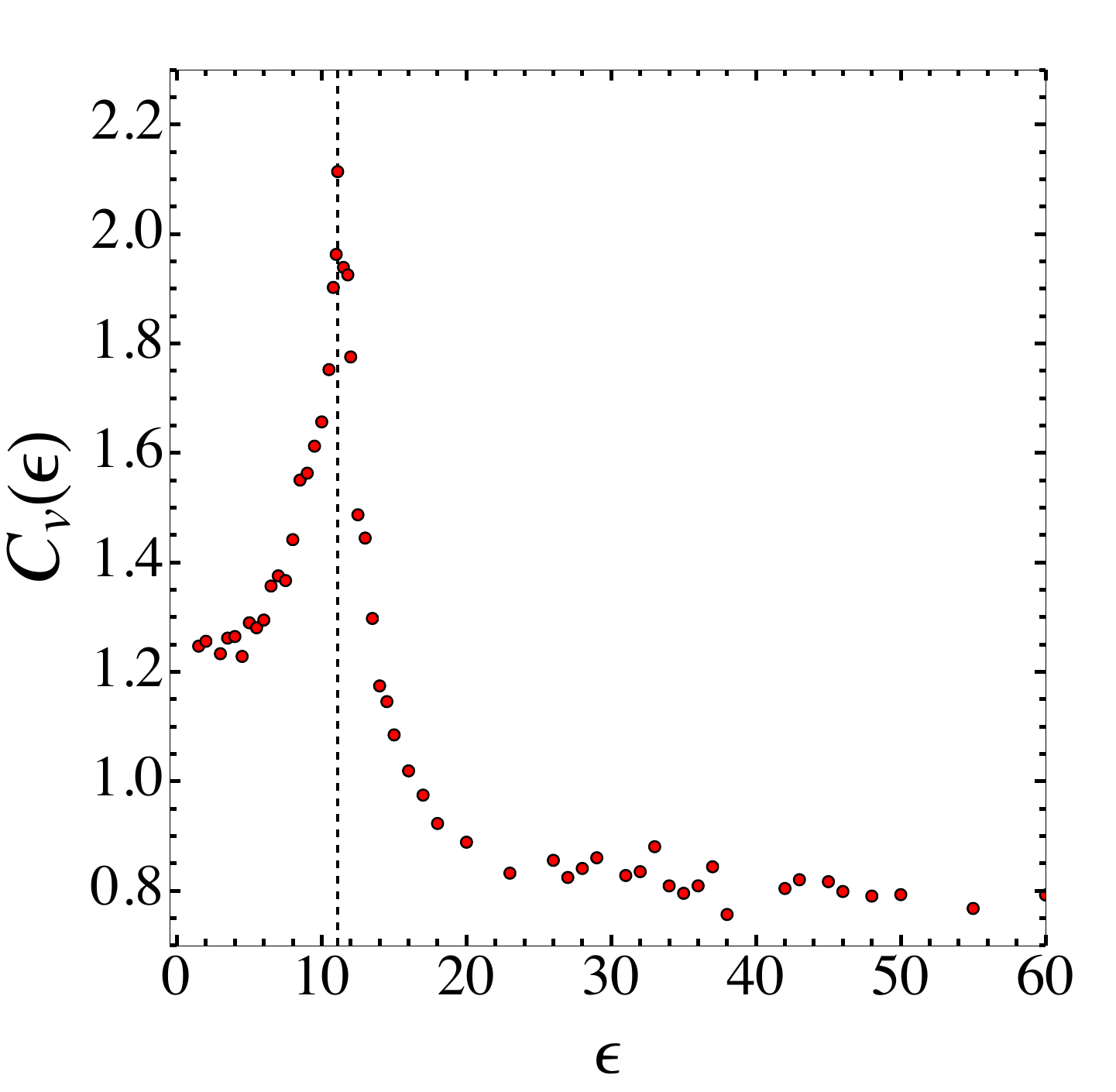}
        \caption{Specific heat.}
        \label{SFig:specific_heat}
    \end{minipage}
    \caption*{Plots of the standard thermodynamics observables for the $\phi^{4}$-model. The vertical dashed lines mark the transition energy point, $\epsilon_t=E_t/N\approx 11.1$.}
    \label{}
\end{figure}

In order to get information about the phase transition and the critical energy point, we numerically computed time averages of the relevant thermodynamics observables such as the caloric curve, magnetization and specific heat. The magnetization is defined by \cite{pettini2007geometry}
\begin{equation}\label{def:magnetization}
    \mathcal{M}:=\sum_{\bm{i}}|\phi_{\bm{i}}|,
\end{equation}
whose average is computed in agreement with Eq. \eqref{def:time_average}. The caloric curve is give, instead, by ($k_B=1$) \cite{pettini2007geometry}
\begin{equation}
    T(E):=\frac{2}{N}\langle K\rangle
\end{equation}
where $\langle K\rangle$ is the time average of the kinetic energy defined through the momenta degrees of freedom $\pi_{i}$.\\
Finally, the specific heat can be computed through averaged of the kinetic energy as follows \cite{pettini2007geometry}
\begin{equation}\label{def:specific_heat}
    C_{v}(E):=\left(N-(N-2)\langle K\rangle\langle K^{-1}\rangle\right)^{-1},
\end{equation}
The numerical results are reported below. The magnetization as defined in Eq. \eqref{def:specific_heat} has been plotted in Fig. \ref{SFig:magnetization}, the caloric curve in Fig. \eqref{SFig:caloric_curve} and, finally, the specific heat defined by \eqref{def:specific_heat} is plotted in Fig. \ref{SFig:specific_heat}.
We remark that the results obtained in this work are in agreement with those published in Ref. \cite{bel2021geometrical}.

\subsection{Geometric observables in $\phi^4$-model}

In order to numerically evaluate $\Omega^{(1)}_{\widetilde{g}}$ and $\Omega^{(2)}_{\widetilde{g}}$, we have to provide an explicit mathematical expression for the Hessian and the gradient of the Hamiltonian function \eqref{def:hamiltonian_phi4}. Let us define $\bm{x}_{\bm{k}}:=(\pi_{\bm{k}},\phi_{\bm{k}})$, then, we have:
\begin{equation}
    \nabla_{\pi_{\bm{k}}}H=\pi_{\bm{k}},\qquad \nabla_{\phi_{\bm{k}}}H=\frac{\lambda}{3!}\phi^{3}_{\bm{k}}+(4J-\mu^2)\phi_{\bm{k}}-J\sum_{\bm{i}\in N\!\!N(\bm{k})}\phi_{\bm{i}},
\end{equation}
then, the components of the gradient of $H$ are given by
\begin{equation}
    \nabla_{\bm{x}_{\bm{k}}} H=(\nabla_{\pi_{\bm{k}}}H,\nabla_{\phi_{\bm{k}}}H).
\end{equation}
The Hessian of $H$ can be obtained by differentiating $\nabla_{\bm{k}}H$, we have
\begin{equation}
    Hess\,H=\left(\begin{matrix}
    1\!\!\!1_{\mathbb{R}^{N}} & \bm{0}_{\mathbb{R}^N}\\
    \bm{0}_{\mathbb{R}^N} & Hess\,V
    \end{matrix}\right),
\end{equation}
where
\begin{equation}
    Hess\,V_{\bm{kj}}=\nabla_{\phi_{\bm{k}}}\nabla_{\phi_{\bm{j}}}V=\left(\frac{\lambda}{2}\phi^{2}_{\bm{j}}+4J-\mu^2\right)\delta_{\bm{kj}}-J\delta_{\bm{j}, N\!\!N(\bm{k})}.
\end{equation}
Moreover, we have
\begin{equation}
\begin{split}
    \nabla\nabla\nabla\,V_{\bm{kjl}}&=\nabla_{\phi_{\bm{k}}}\nabla_{\phi_{\bm{j}}}\nabla_{\phi_{\bm{l}}}V=\lambda\phi_{\bm{j}}\delta_{\bm{kj}}\delta_{\bm{jl}},\\
    \nabla\nabla\nabla\nabla\,V_{\bm{kjli}}&=\nabla_{\phi_{\bm{k}}}\nabla_{\phi_{\bm{j}}}\nabla_{\phi_{\bm{l}}}\nabla_{\phi_{\bm{i}}}V=\lambda\delta_{\bm{ji}}\delta_{\bm{kj}}\delta_{\bm{jl}},\\
    \nabla\nabla\nabla\nabla\nabla\,V_{\bm{kjlim}}&=\nabla_{\phi_{\bm{k}}}\nabla_{\phi_{\bm{j}}}\nabla_{\phi_{\bm{l}}}\nabla_{\phi_{\bm{i}}}\nabla_{\phi_{\bm{m}}}V=0.
\end{split}
\end{equation}
The laplacian of $H$ is given by taking the trace of the Hessian:
\begin{equation}
    \Delta H=N+Tr[Hess\,V]=N+\frac{\lambda}{2}\sum_{\bm{i}}\phi^{2}_{\bm{i}}+N(4J-\mu^2).
\end{equation}

\section{Numerical details for the Ginzburg-Landau-like model}

This Hamiltonian system has been already studied in Ref.~\cite{di2021topology}, we report here just those numerical details which are the strictly necessary for the analysis of the geometric observables.\\

The Ginzburg-Landau-like model is defined by the Hamiltonian 
\begin{equation}\label{def:GL_hamiltonian}
    H_{GL}:=\sum_{i=1}^{n}\frac{p^{2}_{i}}{2}-\frac{\alpha}{2}\sum_{i=1}^{n}(q^{i})^{2}+\frac{\beta}{4}\left(\sum_{i=1}^{n}(q^{i})^{2}\right)^{2},
\end{equation}
where $\alpha,\;\beta\in\mathbb{R}^{+}$.
Depending on the value of $E$, the energy level sets show a change in their topology. In fact, by defining the \emph{order parameter}
\begin{equation}\label{def:order_parameter}
    r^{2}=\sum_{i=1}^{n}(q^{i})^{2},
\end{equation}
together with the total momentum:
\begin{equation}\label{}
    P^{2}=\sum_{i=1}^{n}p_{i}^{2},
\end{equation}
the Hamiltonian function \eqref{def:GL_hamiltonian} rewrites
\begin{equation}\label{def:hamiltonian_function_GL}
    H_{GL}(P,r)=\frac{P^{2}}{2}-\frac{\alpha}{2}r^{2}+\frac{\beta}{4}r^{4}.
\end{equation}
We note that it admits three classes of stationary points given by the condition $\nabla H_{GL}=\bm{0}$, namely:
\begin{equation}
    \frac{\partial H_{GL}}{\partial P}=P=0,\qquad \frac{\partial H_{GL}}{\partial r}=(-\alpha +\beta r^{2})r=0.
\end{equation}
This implies that
\begin{equation}
    r^{\pm}_{m}=\pm\sqrt{\frac{\alpha }{\beta }} \quad {\rm and} \quad P_{m}=0,\qquad r_{M}=0\quad{\rm and}\quad P_{M}=0,
\end{equation}
where the subscript ${m}$ stands for minima and ${M}$ for maximum and, finally, in terms of particle coordinates, we have 
\begin{equation}
    \left(\sum_{i=1}^{n}(q_{\pm}^{i})^{2}\right)^{1/2}=\pm\sqrt{\frac{\alpha }{\beta }}\quad {\rm and}\quad  p_{j}=0,\qquad q^{i}=0\quad {\rm and}\quad  p_{j}=0,\qquad\qquad\forall \; j\in[1,n].
\end{equation}
Therefore, the energy level set are defined by 
\begin{equation}\label{def:energy_level_sets}
    H_{GL}(P,r)=\frac{P}{2}-\frac{\alpha}{2}r^{2}+\frac{\beta}{4}r^{4}=E_{GL}.
\end{equation}
Not all the energy values $E_{GL}$ are associated with an accessible energy level set, in fact:
\begin{equation} 
    \forall \; E_{GL}<H_{GL}(P_{m},r^{\pm}_{m})=-\frac{\alpha ^2}{4 \beta }\implies \Sigma_{E}^{H_{GL}}= \emptyset \ .
\end{equation}
Therefore, the lowest energy value is $E_{GL}^{m}=-\alpha^{2}/4 \beta$ and the accessible level sets are defined by the following range of energies 
\begin{equation}\label{def:energy_range_quartic}
    E\in[-\alpha^{2}/4 \beta,\infty).
\end{equation}
Thus, the energy level sets corresponding to the energy values $H(P_{m},r^{\pm}_{m})\leq E_{GL}< H_{GL}(P_{M},r_{M})$ are homeomorphic to two disjoint hyperspheres:
\begin{equation}
    \Sigma_{E}^{H_{GL}}\simeq \mathbb{S}^{n-1}\cup \mathbb{S}^{n-1},
\end{equation}
As $E_{GL}\to 0$, the energy level set $\Sigma_{0}^{H_{GL}}$ is homeomorphic to the one-point-union of the two previous hyperspheres and this can be written through the wedge sum:
\begin{equation}
    \Sigma_{0}^{H_{GL}}\simeq \mathbb{S}^{n-1}\wedge \mathbb{S}^{n-1}= (\mathbb{S}^{n-1}\cup \mathbb{S}^{n-1})/\sim,
\end{equation}
where $\sim$ is the equivalence relation which identifies a point $x_{1}$ on the first hypersphere with the point $x_{2}$ on the second hypersphere.\\
Finally, for $0\leq E_{GL}< \infty$, the energy level sets are homeomorphic to hyperspheres; hence, we have
\begin{equation}
    \Sigma_{E}^{H_{GL}}\simeq \mathbb{S}^{n-1}.
\end{equation}
We have chosen the following allowed energy subset $I_{GL}=[-\alpha^{2}/4\beta,1]$ and we have sampled energy values from these sets with an energy-step $\Delta E=10^{-4}$.
Then, we have numerically solved the Hamilton's equations with $n=150$ particles adopting a second order bilateral symplectic algorithm \cite{casetti1995efficient}. We set $\alpha=0.5$ and $\beta=0.7$ which implies that $\alpha^{2}/4\beta=0.089$ and an integration time step $\Delta t=10^{-4}$. We note that the closest the energy values to the lowest one the larger the scalar and mean curvatures. In fact, since the energy level sets are always homeomorphic to a sphere or two disjoint spheres, in the $E\to-\alpha^{2}/4\beta$ limit, the level sets reduce their volume till becoming two points. Therefore, in order to avoid floating point overflow, we chose as minimum energy value $E_{GL}^{m}=-0.08$ which is slightly larger than $-\alpha^{2}/4\beta$. For any energy value $E$, random initial conditions have been chosen. Then, we have evaluated $\Omega_{\widetilde{g}}^{(1)}(E)$ along the dynamics at each energy value within $I_{GL}$ through the relation \eqref{def:time_average} and we used it for integrating Eq. (33) numerically obtaining $S_{\widetilde{g}}(E)$. The results are plotted in the main text (see Fig. 1). A least-sensitive inflection point in the entropy function, $S_{\widetilde{g}}$, appears at the transition point $\epsilon_t=0$. In particular, the first-order geometric function $\Omega_{\widetilde{g}}^{(1)}(E)$ has a positive-valued minimum to the left of $\epsilon_t$. The apparent peak in $\Omega_{\widetilde{g}}^{(1)}(E)$ is due to the fact that the GL-model undergoes a PT in correspondence of a critical point. In fact, this is an \emph{extreme} behavior of the function $\beta(E)$ provided by Bachmann (see Fig. 4 in the main text) where the maximum and the minimum points of $\beta$ lie on the black dashed vertical line.

\subsection{Geometric observables in Ginzburg-Landau-like model}

In the Ginzburg-Landau-like model, we define $x_i:=(p_i,q_i)$ and the gradient of the Hamiltonian \eqref{def:GL_hamiltonian} reads:
\begin{equation}
    \nabla_{p_i}H_{GL}=p_i, \qquad \nabla_{q_i}H_{GL}=-\alpha q_i+\beta q_i\left(\sum_{i=1}^{n}q_i^2\right),
\end{equation}
so that 
\begin{equation}
    \nabla_{x_i}H_{GL}=(\nabla_{p_i}H_{GL},\nabla_{q_i}H_{GL}).
\end{equation}
The Hessian matrix is given by 
\begin{equation}
    Hess\,H_{GL}=\left(\begin{matrix}
    1\!\!\!1_{\mathbb{R}^{n}} & \bm{0}_{\mathbb{R}^n}\\
    \bm{0}_{\mathbb{R}^n} & Hess\,V_{GL}
    \end{matrix}\right),
\end{equation}
with
\begin{equation}
    [Hess\,V_{GL}]_{ij}=\left[-\alpha+\beta\left(\sum_{i=1}^{n}q_i^2\right)\right]\delta_{ij}+2\beta\, q_i~q_j
\end{equation}
Moreover, we have
\begin{equation}
\begin{split}
    [\nabla\nabla\nabla\,V_{GL}]_{ijk}&=2\beta\,q_k \delta_{ij}+2\beta (\delta_{ik}q_{j}+\delta_{jk}q_{i}),\\
    [\nabla\nabla\nabla\nabla\,V_{GL}]_{ijkl}&=2\beta\,\delta_{lk} \delta_{ij}+2\beta (\delta_{ik}\delta_{lj}+\delta_{jk}\delta_{il}),\\
    [\nabla\nabla\nabla\nabla\nabla\,V_{GL}]_{ijklm}&=0.
\end{split}
\end{equation}
The laplacian of $H$ is given by taking the trace of the Hessian:
\begin{equation}
    \Delta H_{GL}=n+Tr[Hess\,V_{GL}]=n-\alpha n+\beta\left(\sum_{i=1}^{n}q_i^2\right)(n+2).
\end{equation}

\section{Proof of Universality of the first and second GCFs' energy behaviors}

In this section, we prove that the first and second order GCFs manifest an universal, that is, system-independent energy behavior. In other words, all the Hamiltonian systems which undergo a first (second) order PT à la Bachmann, admit the same qualitative energy behavior in the first (second) order GCF.\\
To reach such a purpose, it is actually sufficient to compute the GCF $\Omega_{\widetilde{g}}^{(1)}$ ($\Omega_{\widetilde{g}}^{(2)}$) for a specific system. Note that this can be easily done numerically. Then, exploiting the Riccati equation (38) and the Cauchy's theorem of existence and uniqueness of the solution of the ordinary differential equations, we can conclude that a different behavior of $\Omega_{\widetilde{g}}^{(1)}$ ($\Omega_{\widetilde{g}}^{(2)}$) cannot reproduce the behavior of $\partial_E^2 S$ predicted by Bachmann.\\
Then, let us start with the Riccati equation for the entropy that we recall below
\begin{equation}\label{eqn:entropy_flow_equation}
    \partial_E^2 S(E)+(\partial_E S(E))^2=\Omega_{\widetilde{g}}^{(2)}.
\end{equation}
then, let us plug the following transformation $S(E)=\log(f(E)/f_0)$ with $f(E_0)=f_0\in\mathbb{R}$ into the equation above. We get
\begin{equation}
\begin{split}\label{def:system_riccati}
    \partial_{E}f(E)= g(E),\qquad
    \partial_{E}g(E)= \Omega_{\widetilde{g}}^{(2)},
\end{split}
\end{equation}
We have now a set of linear first order differential equations which has to be solved for the unknown function $X=(f,g)$. Hence, Eq. \eqref{def:system_riccati} can be rewritten as
\begin{equation}\label{def:compact_riccati}
    \partial_E X(E)=A_{2}(E)\,X(E),
\end{equation}
where
\begin{equation}\label{def:A_matrix}
    A_{2}=\left(\begin{matrix}
    0 & 1\\
    \Omega_{\widetilde{g}}^{(2)} & 0
    \end{matrix}\right)
\end{equation}
By introducing the notation $F(E,X(E)):=A_{2}(E)\,X(E)$, a formal solution of Eq. \eqref{def:compact_riccati} is given by
\begin{equation}
\begin{split}\label{def:solution_R}
    X_{2}(E)=X_{2}(E_0)+\int_{E_0}^{E}F(s,X_{2}(s))~ds,
\end{split}
\end{equation}
Now, let us suppose by contradiction that there exists a $\widehat{\Omega}_{\widetilde{g}}^{(2)}$ with an energy behavior qualitatively different from that of $\Omega_{\widetilde{g}}^{(2)}$ but such that $\partial S^2_E$ admits a negative-valued peak as before. This means that, given the matrix
\begin{equation}\label{def:A_matrix_hat}
    A_{2}=\left(\begin{matrix}
    0 & 1\\
    \widehat{\Omega}_{\widetilde{g}}^{(2)} & 0
    \end{matrix}\right)
\end{equation}
together with the respective Riccati equation
\begin{equation}
    \partial_E \widehat{X}_{2}(E)=\widehat{A}_{2}(E)\widehat{X}_{2}(E).
\end{equation}
we have
\begin{equation}\label{eqn:same_solutions}
    X_{2}=\widehat{X}_{2}\equiv X.
\end{equation}
Hence, let us write the formal solutions
\begin{equation}
\begin{split}
    X_{2}(E)=X_{2}(E_0)+\int_{E_0}^{E}F(s,X_{2}(s))~ds,\qquad
    \widehat{X}_{2}(E)=\widehat{X}_{2}(E_0)+\int_{E_0}^{E}\widehat{F}(s,\widehat{X}_{2}(s))~ds,
\end{split}
\end{equation}
where $\widehat{F}(s,\widehat{X}_{2}(s)):=\widehat{A}_{2}(E)\widehat{X}_{2}(E)$ and let us subtract the first equation with the second one imposing Eq. \eqref{eqn:same_solutions}, we obtain:
\begin{equation}
    \int_{E_0}^{E}(F(s,X_{2}(s))-\widehat{F}(s,X(s)))~ds=0
\end{equation}
which implies $(A_{2}(E)-\widehat{A}_{2}(E))X(E)=0$ then, assuming that $X(E)$ is a non trivial solution, we get $A_{2}(E)=\widehat{A}_{2}(E)$ concluding that $\widehat{\Omega}_{\widetilde{g}}^{(2)}(E)=\Omega_{\widetilde{g}}^{(2)}(E)$, at least, in a neighborhood of the transition point. A similar result for the first-order GCF can easily deduced from the previous one.

%%%%%%%%%%%%%%%%%%%%%%%%%%%


\section*{References}
\begin{thebibliography}{9}



\bibitem{ehrenfest1933phasenumwandlungen}
  P. Ehrenfest, \textit{Phasenumwandlungen im ueblichen und erweiterten Sinn, classifiziert nach den entsprechenden Singularitaeten des thermodynamischen Potentiales}, NV Noord-Hollandsche Uitgevers Maatschappij, 1933.

\bibitem{gross2001microcanonical}
  Gross, Dieter HE, \textit{Microcanonical thermodynamics: phase transitions in" small" systems}, World Scientific, 2001.


\bibitem{gross2005microcanonical}
  Gross, DHE and Kenney, JF, \textit{The microcanonical thermodynamics of finite systems: The microscopic origin of condensation and phase separations, and the conditions for heat flow from lower to higher temperatures}, The Journal of chemical physics, {\bf 122}, 2005.


\bibitem{qi2018classification} Qi, K. and Bachmann, M., \textit{Classification of phase transitions by microcanonical inflection-point analysis}, Physical review letters, {\bf 120}, (2018).


\bibitem{schnabel2011microcanonical} Schnabel, S. and Seaton, D. T. and Landau, D. P. and Bachmann, M., \textit{Microcanonical entropy inflection points: Key to systematic understanding of transitions in finite systems}, Physical Review E {\bf 32}, (2011).



\bibitem{bachmann2014novel} Bachmann, M., \textit{Novel concepts for the systematic statistical analysis of phase transitions in finite systems}, Journal of Physics: Conference Series, IOP Publishing, {\bf 487}, (2014).



\bibitem{bachmann2014thermodynamics} Bachmann, M., \textit{Thermodynamics and statistical mechanics of macromolecular systems}, Cambridge University Press, (2014).


\bibitem{koci2017subphase} Koci, T. and Bachmann, M., \textit{Subphase transitions in first-order aggregation processes}, Physical Review E, {\bf 95}, (2017).



\bibitem{sitarachu2020exact} Sitarachu, K. and Zia, R.K.P. and Bachmann, M., \textit{Exact microcanonical statistical analysis of transition behavior in Ising chains and strips}, Journal of Statistical Mechanics: Theory and Experiment, {\bf 2020}, (2020).



\bibitem{sitarachu2020phase} Sitarachu, K. and Bachmann, M., \textit{Phase transitions in the two-dimensional Ising model from the microcanonical perspective}, Journal of Physics: Conference Series, IOP Publishing, {\bf 1483}, (2020).




% ===================
%.  Topological theory PT

\bibitem{PhysRevLett.79.4361} Caiani, L. and Casetti, L. and Clementi, C. and Pettini, M., \textit{Geometry of Dynamics, Lyapunov Exponents, and Phase Transitions}, Phys. Rev. Lett., {\bf 79}, (1997).


\bibitem{CASETTI2000237} Casetti, L. and Pettini, M. and Cohen, E.G.D. , \textit{Geometric approach to Hamiltonian dynamics and statistical mechanics}, Physics Reports, {\bf 337}, (2000).

\bibitem{caiani1998hamiltonian} Caiani, L. and Casetti, L. and Pettini, M., \textit{Hamiltonian dynamics of the two-dimensional lattice model}, Journal of Physics A: Mathematical and General, {\bf 33137}, (1998).


\bibitem{caiani1998geometry} Caiani, L. and Casetti, L. and Clementi, C. and Pettini, G. and Pettini, M. and Gatto, R., \textit{Geometry of dynamics and phase transitions in classical lattice $\varphi$ 4 theories}, Physical Review E, {\bf 57}, (1998).

%============================== QUI

\bibitem{franzosi1999topological} Franzosi, Roberto and Casetti, Lapo and Spinelli, Lionel and Pettini, Marco, \textit{Topological aspects of geometrical signatures of phase transitions}, Physical Review E, {\bf 60}, (1999).


\bibitem{pettini2007geometry} Pettini, Marco, \textit{Geometry and topology in Hamiltonian dynamics and statistical mechanics}, Springer Science \& Business Media (2007).



\bibitem{franzosi2004theorem} Franzosi, Roberto and Pettini, Marco, \textit{Theorem on the origin of phase transitions}, Physical Review Letters, {\bf 92}, (2004).


\bibitem{bel2021geometrical} Bel-Hadj-Aissa, Ghofrane and Gori, Matteo and Franzosi, Roberto and Pettini, Marco, \textit{Geometrical and topological study of the Kosterlitz--Thouless phase transition in the XY model in two dimensions}, Journal of Statistical Mechanics: Theory and Experiment, {\bf 2021}, (2021).


\bibitem{rousset2010free} Rousset, Mathias and Stoltz, Gabriel and Lelievre, Tony, \textit{Free Energy Computations: A Mathematical Perspective}, World Scientific, (2010).




\bibitem{gori2018topological} Gori, Matteo and Franzosi, Roberto and Pettini, Marco, \textit{Topological origin of phase transitions in the absence of critical points of the energy landscape}, Journal of Statistical Mechanics: Theory and Experiment, {\bf 2018}, (2018).





\bibitem{chen2014total} Chen, Bang-Yen, \textit{Total mean curvature and submanifolds of finite type}, World Scientific Publishing Company, {\bf 27}, (2014).


\bibitem{pettini2019origin} Pettini, Giulio and Gori, Matteo and Franzosi, Roberto and Clementi, Cecilia and Pettini, Marco, \textit{On the origin of phase transitions in the absence of symmetry-breaking}, Physica A: Statistical Mechanics and its Applications, {\bf 516}, (2019).



% Criticism
\bibitem{mehta2012energy} Mehta, Dhagash and Hauenstein, Jonathan D and Kastner, Michael, \textit{Energy-landscape analysis of the two-dimensional nearest-neighbor $\varphi$ 4 model}, Physical Review E, {\bf 85}, (2012).


\bibitem{kastner2011phase} Kastner, Michael and Mehta, Dhagash, \textit{Phase transitions detached from stationary points of the energy landscape}, Physical Review Letters, {\bf 107}, (2011).



\bibitem{rugh1997dynamical} Rugh, Hans Henrik, \textit{Dynamical approach to temperature}, Physical Review Letters, {\bf 78}, (1997).


\bibitem{rugh1998geometric} Rugh, Hans Henrik, \textit{A geometric, dynamical approach to thermodynamics}, Journal of Physics A: Mathematical and General, {\bf 31}, (1998).




\bibitem{rugh2001microthermodynamic} Rugh, Hans Henrik, \textit{Microthermodynamic formalism}, Physical Review E, {\bf 64}, (2001).





\bibitem{bel2020geometrical} Bel-Hadj-Aissa, Ghofrane and Gori, Matteo and Penna, Vittorio and Pettini, Giulio and Franzosi, Roberto, \textit{Geometrical aspects in the analysis of microcanonical phase-transitions}, Entropy, {\bf 22}, (2020).



\bibitem{franzosi2018microcanonical} Franzosi, Roberto, \textit{Microcanonical entropy for classical systems}, Physica A: Statistical Mechanics and its Applications, {\bf 494}, (2018).
% definition of entropy Franzosi


\bibitem{franzosi2019microcanonical} Franzosi, Roberto and Casetti, Lapo and Spinelli, Lionel and Pettini, Marco, \textit{A microcanonical entropy correcting finite-size effects in small systems}, Journal of Statistical Mechanics: Theory and Experiment, {\bf 2019}, (2019).



\bibitem{di2021topology} Di Cairano, Loris and Gori, Matteo and Pettini, Marco, \textit{Topology and Phase Transitions: A First Analytical Step towards the Definition of Sufficient Conditions}, Entropy, {\bf 23}, (2021).



% ====================================

\bibitem{zhou2013simple} Zhou, Yajun, \textit{A simple formula for scalar curvature of level sets in Euclidean spaces}, arXiv preprint, arXiv:1301.2202, (2013).




\bibitem{gromov2019four} Gromov, Misha, \textit{Four lectures on scalar curvature}, arXiv preprint, arXiv:1908.10612, (2019).


\bibitem{hirsch2012differential} Hirsch, Morris W, \textit{Differential topology}, Springer Science \& Business Media, {\bf 33}, (2012).


\bibitem{franzosi2000topology} Franzosi, Roberto and Pettini, Marco and Spinelli, Lionel, \textit{Topology and phase transitions: Paradigmatic evidence}, Physical Review Letters, {\bf 84}, (2000).



\bibitem{spivak1975comprehensive} Spivak, Michael, \textit{A comprehensive introduction to differential geometry}, Publish or Perish, Incorporated, {\bf 5}, (1975).





\bibitem{gori_thesis} Gori, Matteo, \textit{Phase Transitions Theory and applications to Biophysics}, PhD Thesis, (2016).



\bibitem{petersen2006riemannian} Petersen, Peter, \textit{Riemannian geometry}, Springer, {\bf 171}, (2006).



\bibitem{casetti1995efficient} Casetti, Lapo, \textit{Efficient symplectic algorithms for numerical simulations of Hamiltonian flows}, Physica scripta, {\bf 51}, (1995).


\bibitem{bittanti2012riccati} Bittanti, Sergio and Laub, Alan J and Willems, Jan C, \textit{The Riccati Equation}, Springer Science \& Business Media, (2012).





% discontinuity entropy derivative finite size

\bibitem{nerattini2013exploring} Nerattini, Rachele and Kastner, Michael and Mehta, Dhagash and Casetti, Lapo, \textit{Exploring the energy landscape of X Y models}, Physical Review E, {\bf 87}, (2013).



\bibitem{kastner2006mean} Kastner, Michael and Schnetz, Oliver, \textit{On the mean-field spherical model}, Journal of statistical physics, {\bf 122}, (2006).


\bibitem{kastner2008phase} Kastner, Michael, \textit{Phase transitions and configuration space topology}, Reviews of Modern Physics, {\bf 80}, (2008).



\bibitem{casetti2006nonanalyticities} Casetti, Lapo and Kastner, Michael, \textit{Nonanalyticities of entropy functions of finite and infinite systems}, Physical Review letters, {\bf 97}, (2006).


\bibitem{hilbert2006nonanalytic} Hilbert, Stefan and Dunkel, J{\"o}rn, \textit{Nonanalytic microscopic phase transitions and temperature oscillations in the microcanonical ensemble: An exactly solvable one-dimensional model for evaporation}, Physical Review E, {\bf 74}, (2006).


\bibitem{dunkel2006phase} Dunkel, J{\"o}rn and Hilbert, Stefan, \textit{Phase transitions in small systems: Microcanonical vs. canonical ensembles}, Physica A: Statistical Mechanics and its Applications, {\bf 370}, (2006).



\bibitem{angelani2005topology} Angelani, Luca and Casetti, Lapo and Pettini, Marco and Ruocco, Giancarlo and Zamponi, Francesco, \textit{Topology and phase transitions: From an exactly solvable model to a relation between topology and thermodynamics}, Physical Review E, {\bf 71}, (2005).


\bibitem{casetti2009kinetic} Casetti, Lapo and Kastner, Michael and Nerattini, Rachele, \textit{Kinetic energy and microcanonical nonanalyticities in finite and infinite systems}, Journal of Statistical Mechanics: Theory and Experiment, {\bf 2009}, (2009).


\bibitem{matsumoto2002introduction} Matsumoto, Yukio, \textit{An introduction to Morse theory}, American Mathematical Soc., {\bf 208}, (2002).


\bibitem{stein1963morse} Stein, Elias and Milnor, John Willard and Spivak, Michael and Wells, R and Wells, Robert and Mather, John N, \textit{Morse theory}, Princeton University Press, (1963).


\bibitem{di2019coherent} Di Cairano, Loris and Gori, Matteo and Pettini, Marco, \textit{Coherent Riemannian-geometric description of Hamiltonian order and chaos with Jacobi metric}, Chaos: An Interdisciplinary Journal of Nonlinear Science, {\bf 29}, (2019).


\bibitem{di2021hamiltonian} Di Cairano, Loris and Gori, Matteo and Pettini, Giulio and Pettini, Marco, \textit{Hamiltonian chaos and differential geometry of configuration space--time}, Physica D: Nonlinear Phenomena, Elsevier, {\bf 422}, (2021).


\bibitem{arnol2013mathematical} Arnol'd, Vladimir Igorevich, \textit{Mathematical methods of classical mechanics}, Springer Science \& Business Media, {\bf 60}, (2013).


\bibitem{casetti1996riemannian} Casetti, Lapo and Clementi, Cecilia and Pettini, Marco, \textit{Riemannian theory of Hamiltonian chaos and Lyapunov exponents}, Physical Review E, {\bf 54}, (1996).


\bibitem{eisenhart1928dynamical} Eisenhart, Luther Pfahler, \textit{Dynamical trajectories and geodesics}, Annals of Mathematics, JSTOR, {\bf 60}, (1928).


\bibitem{pettini1993geometrical} Pettini, Marco, \textit{Geometrical hints for a nonperturbative approach to Hamiltonian dynamics}, Physical Review E, {\bf 47}, (1993).



\bibitem{casetti1993analytic} Casetti, Lapo and Pettini, Marco, \textit{Analytic computation of the strong stochasticity threshold in Hamiltonian dynamics using Riemannian geometry}, Physical Review E, {\bf 48}, (1993).


\bibitem{kastner2008phasevanishing} Kastner, Michael and Schnetz, Oliver, \textit{Phase transitions induced by saddle points of vanishing curvature}, Physical Review Letters, {\bf 100}, (2008).



\bibitem{kastner2008nonanalyticities} Kastner, Michael and Schnetz, Oliver and Schreiber, Steffen, \textit{Nonanalyticities of the entropy induced by saddle points of the potential energy landscape}, Journal of Statistical Mechanics: Theory and Experiment, {\bf 2008}, (2008).




\end{thebibliography}
\end{document}